# Quantifying Influence and Information Transfer in a Modified Vicsek Model with Non-reciprocal Interactions

Jiahuan Pang[1] and Wendong Wang[1,2,3] *

[1]Global College, Shanghai Jiao Tong University, Shanghai 200240, China.

[2]National Key Laboratory of Advanced Micro and Nano Manufacture Technology, Shanghai Jiao Tong University, Shanghai, 200240, China

[3]State Key Laboratory of Mechanical System and Vibration, Shanghai Jiao Tong University, Shanghai, 200240, China

**ABSTRACT**. Understanding information transfer among individuals is fundamental to revealing the collective dynamics of complex systems. Information transfer has been quantified using various information-theoretic tools and assigned the concept of influence. However, information-theoretic measures are inherently statistical, not causal, and influence in the context of causal inference implies a causal relation, so equating influence with information transfer creates conceptual confusion and interpretational challenges. Here, we introduce an influence-based Vicsek model with non-reciprocal interactions to distinguish influence from information transfer and examine their relationship. At the pairwise level, for fixed noise strengths, influence and transfer entropy exhibit quasi-linear relations; for fixed interaction weights, influence and transfer entropy exhibit nonlinear relations because noise on influencers enhances information transfer, whereas noise on followers suppresses the overall information transfer. At the collective level, we find that both influence and normalized transfer entropy form two-branched relations that clearly identify the transition points across three distinct phase transitions. These transition points reveal a different aspect of the collective dynamics not captured by classical order parameters: phase transitions are associated with changes in the relative importance of influencers' presents or followers' presents on followers' futures. Finally, we use our model to assess partial information decomposition methods and identify two methods most suitable for analyzing our system, one based on pointwise surprisal changes and the other on secret key agreement. Our work is a first step in distinguishing the concept of influence from information transfer in a physical model system, provides a concrete testbed for methods emerging from the growing field of information theory-based causal inference, and offers new insights into the dynamics of complex systems.

## I. INTRODUCTION.

The flow of information is crucial to the dynamics of complex systems [1–4], such as biological reaction networks [5], social networks [6], neural networks [7], bird flocks [8], fish schools [9], and evolving systems [10]. These complex systems contain many interacting units, and quantifying the information transfer among these units helps identify leaders and followers, as well as attributing causes and effects. Information transfer has been calculated primarily using information-theoretic tools, such as mutual information (MI) and transfer entropy (TE) [1,11]. Recently, the validity of these tools has been called into question [4,12], which spurred more efforts to devise new information-theoretical tools [13–17]. However, the proliferation of information-theoretic tools creates challenges for tool selection and for the interpretation of results. Moreover, all of these tools are based on multivariate joint distributions and are therefore statistical in nature [4].

Influence is the term that has been assigned to the quantities that the information-theoretical tools calculate [18,19]. This usage stems, perhaps, from the word's common-sense meaning. Following this usage, one assumes that more information transferred means higher influence, and vice versa. This assumption is never explicitly justified and warrants some close examination. Consider a flock of birds [Fig. 1]: We observe their headings and trajectories, from which we calculate information transfer using information-theoretical tools. These tools, however, produce different values of information transfer, which creates confusion in interpretation. On the other hand, influence, we argue, is a part of individual birds' mental decision-making processes and is their response to other birds' behaviors. Therefore, influence is inherently internal to individual birds and should be regarded as a different concept from the information transfer calculated from birds' headings and trajectories, which are externally observed and measured. Distinguishing these two concepts is the first step in creating an independent test for various information-theoretical tools. Because experimental measurements of influence from individual birds' brains are difficult, a physicist's approach would be to construct a model that allows direct calculation of influence, ideally analytically.

*Contact author: wendong.wang@sjtu.edu.cn

More formally, influence, in the field of causal inference [20–22], refers to causal influence or physical influence [22]. It means that changing one variable changes another variable in the analysis of a cause-and-effect relationship [22]. In other words, influence implies a causal relation: that $X$ influences $Y$ means that changing $X$ will change $Y$. Quantifying influence using information-theoretic tools is challenging due to the statistical nature of these tools [4]. An alternative method for quantifying influence is to construct a physical model that allows direct calculation of influence. This method will not only bring conceptual clarity but also serve as a basis for comparing influence and information transfer quantitatively.

To construct an influence-based model, we start with the Vicsek model. Vicsek model is the first and simplest model of flocking. Although other more complicated and realistic models, such as the inertial spin model [23] and the allocentric model, exist [24], the simplicity of the Vicsek model renders it still a powerful tool for fundamental investigations. We adopt the Vicsek model because it allows us to calculate pairwise influence analytically (after some modifications) and because it produces non-equilibrium phase transitions whose nature has been under debate and is of interest across many areas of inquiry. One such area is non-reciprocal interactions. Introducing non-reciprocal interactions into the Vicsek model produces a richer variety of non-equilibrium phase transitions, whose pairwise and collective influences and information transfer have not been analyzed before.

Non-reciprocity, by breaking action-reaction symmetry at the microscopic level, forbids the description of the system by a global energy function, placing it firmly within the realm of non-equilibrium dynamics [25–29]. One key implication of non-reciprocity is that it can give rise to special non-equilibrium phases, such as chiral motion, swap oscillations, and traveling patterns [25]. The study of these transitions driven by non-reciprocity not only enriches our fundamental understanding of critical phenomena [26] but also offers guidance for designing highly robust functional states in synthetic and biological systems, where directed information flow and broken symmetry are inherent [30–32].

Here, we use the Vicsek model [33–35] as an example and propose an intuitive and quantifiable definition of influence to investigate the quantitative relations between influence and information transfer [Fig. 1]. We modify the original Vicsek model [33] to allow for non-reciprocal interactions and the direct calculation of pairwise influence. By constructing appropriate averages of pairwise influences, we establish their relations with TE and normalized TE, both at the pairwise and collective levels. Moreover, we analyze the dual effects of noise: noise acting on influencers (the source of influence) can enhance information transfer, whereas noise acting on followers (the receivers) suppresses it. Finally, as an application, we select three representative partial information decomposition (PID) methods and evaluate their performance using our model. Although no method gives ideal results, we find that two PID methods are appropriate for analyzing the phase transitions in the modified Vicsek model: one based on pointwise changes in surprisal [36] and another derived from the secret key agreement framework [16].

The rest of the paper is organized as follows. In Sec. II, we briefly review relevant information-theoretic quantities to make our discussion self-contained. In Sec. III, we introduce our influence-based Vicsek model with non-reciprocal interactions and analyze its three distinct phase transitions using traditional order parameters. In Sec. IV, we analyze the relations between influence and information transfer in pairwise interactions. In Sec. V, we analyze the relations between influence and information transfer in collective interactions and test representative PID methods. We conclude in Sec. VI.

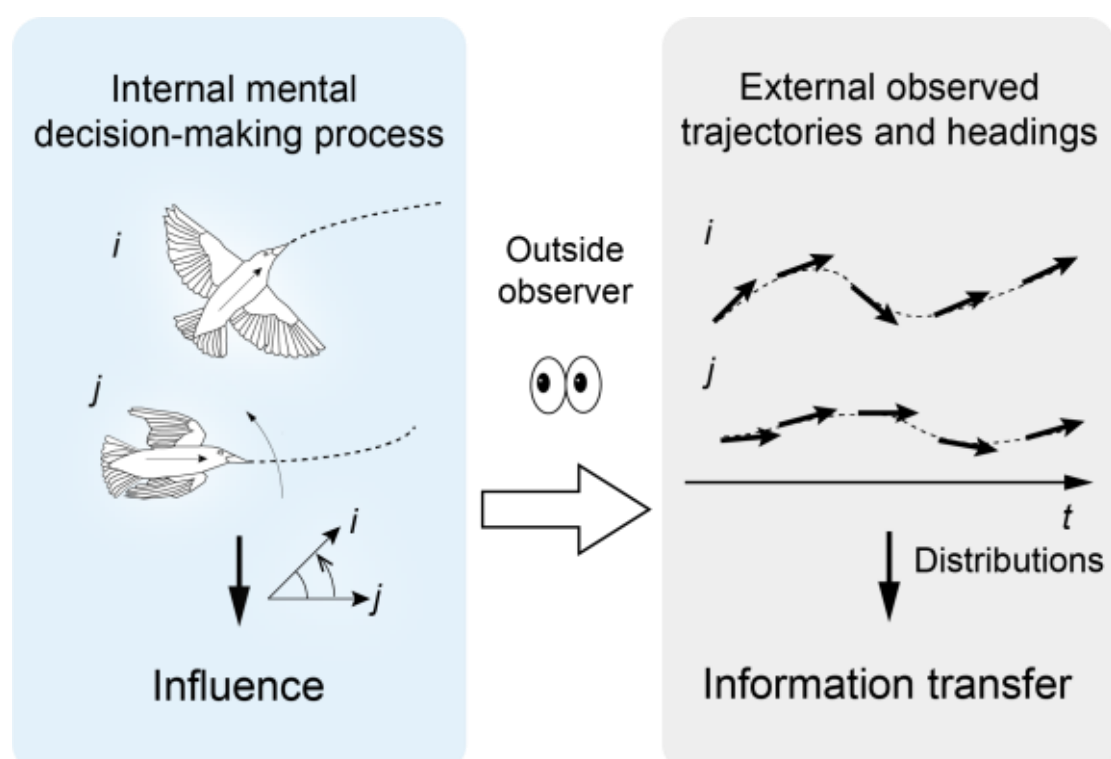

FIG. 1. Relationship between influence and information transfer. The influence by bird $i$ on bird $j$ is a part of bird $j$'s mental decision-making process, so it is inherently internal to each individual bird. In contrast, information transfer is external because it is computed from the birds' trajectories and headings, which can be measured by an outside observer.

## II. BRIEF REVIEW OF RELEVANT INFORMATION THEORETICAL QUANTITIES

This section gives a brief review of the information-theoretical quantities used in this study to quantify information transfer.

### A. Foundational concepts

We begin by introducing foundational concepts in information theory. We use capital letters to denote random variables and lowercase letters to denote their values [37]. Consider a discrete random variable $X$ with alphabet $\mathcal{X}$ and probability mass function $p(x) = \Pr\{X = x\}$. The entropy of $X$ is defined as

$$H(X) = -\sum_{x \in \mathcal{X}} p(x) \log p(x). \quad (1)$$

The base of the log determines the unit of entropy; when the base is 2, entropy is expressed in bits. $H(X)$ quantifies the average uncertainty about $X$ and represents the minimum average number of bits required to encode independent samples of $X$.

Similarly, consider a pair of discrete random variables $(X, Y)$ with a joint distribution $p(x, y)$, the joint entropy $H(X, Y)$ is defined as

$$H(X,Y) = -\sum_{x \in \mathcal{X}} \sum_{y \in \mathcal{Y}} p(x,y) \log p(x,y). \quad (2)$$

$H(X, Y)$ quantifies the average uncertainty about the pair $(X, Y)$, and it is the minimum average number of bits to encode independent joint draws of $(X, Y)$.

Consider again a pair of discrete random variables $(X, Y)$ with a joint distribution $p(x, y)$, the conditional entropy $H(Y|X)$ is defined as

$$H(Y|X) = -\sum_{x \in \mathcal{X}} \sum_{y \in \mathcal{Y}} p(x,y) \log p(y|x), \quad (3)$$

where $p(y|x) = p(x,y)/p(x)$ is the conditional probability. Because $H(Y|X) = H(X,Y) - H(X)$, $H(Y|X)$ quantifies the average remaining uncertainty about $Y$ given the knowledge of $X$.

The relative entropy or Kullback-Leibler distance between two probability mass functions $p(x)$ and $q(x)$ is defined as

$$D(p||q) = \sum_{x \in \mathcal{X}} p(x) \log \frac{p(x)}{q(x)}. \quad (4)$$

Note that $D(p||q) = -\sum p(x) \log q(x) - H(X)$. The term $-\sum p(x) \log q(x)$ quantifies the average uncertainty about $X$ using the wrong distribution $q(x)$ when the true distribution is $p(x)$. Hence, the relative entropy quantifies the average excess uncertainty about $X$ using the wrong distribution $q(x)$ when the true distribution is $p(x)$.

The mutual information (MI) between two random variables $X$ and $Y$ is the relative entropy between the joint distribution $p(x, y)$ and the product of individual distributions $p(x)p(y)$:

$$I(X;Y) = \sum_{x \in \mathcal{X}} \sum_{y \in \mathcal{Y}} p(x,y) \log \frac{p(x,y)}{p(x)p(y)}. \quad (5)$$

MI is symmetric with respect to $X$ and $Y$ and quantifies the amount of information shared between them. It can also be expressed as $I(X;Y) = D\big(p(x,y)||p(x)p(y)\big)$, representing the pair's deviation from independence. Alternatively, $I(X;Y) = H(X) - H(X|Y) = H(Y) - H(Y|X)$, indicating the reduction in uncertainty about one variable given the knowledge of the other.

### B. Information-theoretical measures for information transfer

Building on the above foundational concepts, we now introduce information-theoretical quantities used to measure information transfer between two stochastic processes. Consider two stationary Markov processes $\{\dots, X(t-\Delta t), X(t), X(t+\Delta t), \dots\}$ and $\{\dots, Y(t-\Delta t), Y(t), Y(t+\Delta t), \dots\}$, where $t$ represents the time, and $\Delta t$ represents the discrete time step. We are interested in quantifying how much information is transferred from one process to another process per time step on average. The Markov property means that the processes' future (at $t + \Delta t$) depends only on their present (at $t$) and not on their past (at $t - \Delta t$, $t - 2\Delta t$, …), so it simplifies the task tremendously. We now introduce information-theoretical quantities to measure the information flow from $X$ to $Y$.

Time-delayed mutual information (TDMI), denoted as $TDMI_{X\to Y}$, is defined as

$$\begin{aligned} TDMI_{X\to Y} &= I\big(X(t); Y(t+\Delta t)\big) \\ &= \sum_{x(t)} \sum_{y(t+\Delta t)} p\big(x(t), y(t+\Delta t)\big) \\ &\quad \times \log \frac{p\big(x(t), y(t+\Delta t)\big)}{p\big(x(t)\big)p\big(y(t+\Delta t)\big)}. \end{aligned} \quad (6)$$

Although MI is symmetric and non-directional, TDMI must be interpreted as directional because it relates the present state of $X$ to the future state of $Y$. Information is transferred only from the present to the future, not vice versa. TDMI quantifies the reduction in uncertainty about $Y$'s future given the knowledge of $X$'s present. However, TDMI cannot distinguish actual information transferred between the two processes from the shared information arising from common history or common input sources; for example, if both $X$ and $Y$ receive input from a common source, TDMI may attribute shared information to information transfer erroneously [16].

Transfer entropy (TE) is introduced to overcome the drawback of TDMI [11] and has become the most well-established measure of information transfer [4].

It is in the form of conditional mutual information and is defined as

$$TE_{X\to Y} = I\big(Y(t+\Delta t); X(t)|Y(t)\big)$$
$$= \sum_{y(t+\Delta t)} \sum_{x(t)} \sum_{y(t)} p\big(y(t+\Delta t), x(t), y(t)\big) \times \log \frac{p\big(y(t+\Delta t), x(t)|y(t)\big)}{p\big(x(t)|y(t)\big) p\big(y(t+\Delta t)|y(t)\big)}. \quad (7)$$

Note that $TE_{X\to Y} = H\big(Y(t+\Delta t)|Y(t)\big) - H\big(Y(t+\Delta t)|X(t), Y(t)\big)$, so $TE_{X\to Y}$ quantifies the reduction in uncertainty about $Y(t+\Delta t)$ by the knowledge of $X(t)$, given that we also know $Y(t)$. Note also that $TE_{X\to Y} = I\big(Y(t+\Delta t); X(t), Y(t)\big) - I\big(Y(t+\Delta t); Y(t)\big)$, so $TE_{X\to Y}$ is the total information shared between $Y(t+\Delta t)$ and the pair $\big(X(t), Y(t)\big)$ minus the information shared between $Y(t+\Delta t)$ and $Y(t)$. TE has been adopted as the standard tool for measuring information transfer in complex networks appeared in many areas of science and engineering. Yet it can, however, include information not from either source individually but from two sources simultaneously. For example, reference [12] shows that in a binary system of Bernoulli trials where $Y(t+\Delta t) = X(t)\,\mathrm{XOR}\,Y(t)$, $TE_{X\to Y} > TDMI_{X\to Y}$, which contradicts the intuition that TE reduces the uncertainty about $Y(t+\Delta t)$ by conditioning on $X(t)$.

To understand and resolve the seemingly paradoxical behavior of TE, researchers resort to partial information decomposition (PID) [13,38]. According to PID, the total mutual information $I\big(X(t), Y(t); Y(t+\Delta t)\big)$ between the pair of source variables $\{X(t), Y(t)\}$ and the target variable $Y(t+\Delta t)$ is decomposed into four components [Fig. 2]:

$$I\big(X(t), Y(t); Y(t+\Delta t)\big) = Unq(X) + Unq(Y) + Shd + Syn, \quad (8)$$

where $Unq(X)$ is the unique (intrinsic) information from $X(t)$ to $Y(t+\Delta t)$; $Unq(Y)$ is the unique (intrinsic) information from $Y(t)$ to $Y(t+\Delta t)$; $Shd$ is the shared (redundant) information from both $X(t)$ and $Y(t)$ to $Y(t+\Delta t)$; $Syn$ is the synergistic information from neither $X(t)$ nor $Y(t)$ alone, but from them combined. Moreover, the mutual information between $Y(t+\Delta t)$ and $X(t)$ or $Y(t)$ alone is a sum of their respective unique information and shared (redundant) information:

$$I\big(X(t); Y(t+\Delta t)\big) = Unq(X) + Shd, \quad (9)$$

$$I\big(Y(t); Y(t+\Delta t)\big) = Unq(Y) + Shd. \quad (10)$$

There are three equations with four unknowns, so one additional equation is needed to solve the equation set. There is no generally agreed method to construct the 4th equation. Williams and Beer, the original proposers of PID, constructed a measure of shared information to complete the set [13]. James et al. proposed a method to construct unique information from intrinsic mutual information, which is the upper bound of secret key agreement rate in cryptography [16]. Other selected methods of PID are briefly described in Appendix A.

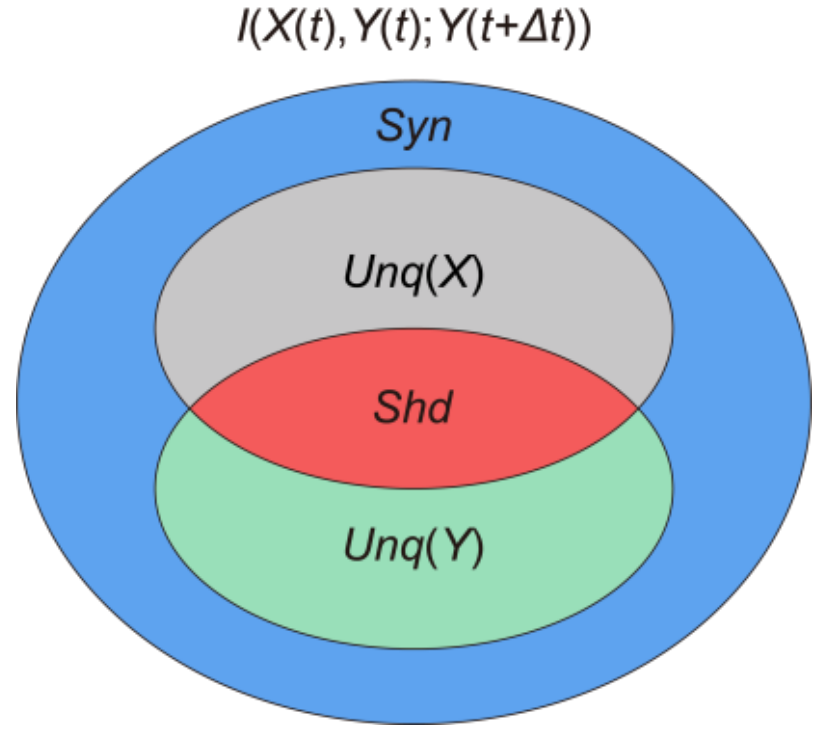

FIG. 2. The structure of partial information decomposition for two source variables and one target variable. $X(t)$ and $Y(t)$ are the two source variables, and $Y(t+\Delta t)$ is the target variable. The grey or the green parts represent the unique information (also called the intrinsic information) from $X(t)$ or $Y(t)$, respectively, to $Y(t+\Delta t)$. The red part represents the shared information (also called the redundant information) from both $X(t)$ and $Y(t)$ to $Y(t+\Delta t)$. The blue part represents the synergistic information from neither $X(t)$ nor $Y(t)$ alone, but from them combined. The entire area of all colored parts represents the total information from $X(t)$ and $Y(t)$ to $Y(t+\Delta t)$ and is the total mutual information $I\big(X(t), Y(t); Y(t+\Delta t)\big)$ shared between the pair of the source variables $\{X(t), Y(t)\}$ and the target variable $Y(t+\Delta t)$. $TE_{X\to Y}$ is the sum of grey and blue parts. $TDMI_{X\to Y}$ is the sum of grey and red parts.

Within the context of PID, $TE_{X\to Y}$ is the sum of the unique information from $X(t)$ and the synergistic information:

$$TE_{X\to Y} = Unq(X) + Syn. \quad (11)$$

Similarly, $TDMI_{X\to Y}$ is the sum of the unique information from $X(t)$ and the shared information:

$$TDMI_{X\to Y} = Unq(X) + Shd. \quad (12)$$

Hence, according to PID, the conditioning on $Y(t)$ in $TE_{X\to Y}$ removes the shared information but introduces the synergistic information, so in systems where synergistic information is larger than shared information, $TE_{X\to Y}$ is greater than $TDMI_{X\to Y}$.

We discretize probability distributions into 8 equal-sized bins using equidistant binning and compute all information-theoretical quantities with the dit package [39]. We adopt this discretization because most PID methods operate on discrete probability mass functions, whereas many alternative estimators for transfer entropy [40–43] are designed to approximate continuous probability density functions.

## III. INFLUENCE-BASED VICSEK MODEL WITH NON-RECIPROCAL INTERACTIONS

We propose a quantifiable definition of influence grounded in three intuitive criteria, which we introduce within the context of the Vicsek model—an archetypal example of a collective system [44]. In the original Vicsek model [33], the future orientation $\theta_i(t+\Delta t)$ of particle $i$ is the summation of the average of its neighbors' present orientations $\langle\theta(t)\rangle$, plus a stochastic noise term $\beta_i(t)$, i.e., $\theta_i(t+\Delta t) = \langle\theta(t)\rangle + \beta_i(t)$. Our first intuitive criterion is that a larger orientational difference between two particles should correspond to a greater pairwise influence within the framework of the Vicsek model. This intuition suggests that the influence by particle $j$ on particle $i$, to a first order approximation, should be proportional to the difference in orientation $\theta_j(t) - \theta_i(t)$. However, directly extracting such a term from the original model is nontrivial because it employs a nonlinear trigonometric operation, $\arctan\left[\langle\sin(\theta(t))\rangle / \langle\cos(\theta(t))\rangle\right]$, to calculate the average of neighbors' orientations $\langle\theta(t)\rangle$. To allow an explicit calculation of pairwise influence, therefore, we need to preserve the orientational difference $\theta_j(t) - \theta_i(t)$ in the averaging operation. In addition, we want to allow asymmetry in the pairwise interactions, such that the influence from an influencer to a follower is different from the influence from a follower to an influencer, specifically to allow for non-reciprocal interactions (We adopt the terms influencer and follower throughout this manuscript.). We therefore introduce a directional interaction weight $w_{j\to i}$ and define a core interaction term as $w_{j\to i}F\left(\theta_j(t) - \theta_i(t)\right)$. Here, $w_{j\to i}$ represents the interaction strength particle $j$ exerts on particle $i$, and $F(x)$ is a wrapping function that confines angular differences to the interval $(-\pi, \pi]$:

$$F(x) = \begin{cases} x\%2\pi, & x\%2\pi \le \pi, \\ x\%2\pi - 2\pi, & x\%2\pi > \pi, \end{cases} \tag{13}$$

where % represents the modulo operation. The introduction of the $w_{j\to i}$ permits the definition of non-reciprocal interactions. A negative weight ($w_{j\to i} < 0$) indicates that particle $i$ tends to align opposite to particle $j$. In this case, the influence should be proportional to the core interaction term, $|w_{j\to i}|F\left(\theta_j(t) + \pi - \theta_i(t)\right)$, where $\theta_j + \pi$ represents the direction opposite to $\theta_j$. Intuitively, this core interaction term is minimized when $\theta_i$ is already aligned opposite to $\theta_j$. Finally, to capture the intuition that the influence of any single neighbor diminishes as the total number of neighbors increases, we normalize the core interaction term by the sum of the absolute interaction weights of all neighbors. Combining these criteria, we define the pairwise instantaneous influence of particle $j$ on particle $i$ as:

$$A_{j\to i}(t) = \begin{cases} \dfrac{|w_{j\to i}|F\left(\theta_j(t) + \pi - \theta_i(t)\right)s_{ij}(t)}{\sum_j |w_{j\to i}|s_{ij}(t)}, & w_{j\to i} < 0, \\ \dfrac{|w_{j\to i}|F\left(\theta_j(t) - \theta_i(t)\right)s_{ij}(t)}{\sum_j |w_{j\to i}|s_{ij}(t)}, & w_{j\to i} \ge 0. \end{cases} \tag{14}$$

Here, $s_{ij}(t) = \Theta\left(R - r_{ij}(t)\right)$ is the neighborhood indicator function, with $\Theta$ being the Heaviside step function, $R$ being the neighborhood cutoff radius, and $r_{ij}(t)$ the distance between particles $i$ and $j$.

The influence of all neighbors on particle $i$ at time $t$ is the sum of pairwise influences:

$$A_i(t) = \sum_j A_{j\to i}(t). \tag{15}$$

$A_i(t)$ can be regarded as the weighted average influence by all neighboring particles on particle $i$ at time $t$. Consequently, the orientation of particle $i$ at the time step $t + \Delta t$ can be written as the sum of its orientation at the current time step $\theta_i(t)$, the weighted average influence $A_i(t)$, and noise $\beta_i(t)$. We constrain the sum to be within $(-\pi, \pi]$ and obtain the expression of the future orientation of particle $i$:

$$\theta_i(t + \Delta t) = F\left(\theta_i(t) + A_i(t) + \beta_i(t)\right). \tag{16}$$

The noise is uniformly distributed in $[-\eta/2, \eta/2]$, and $\eta \in [0, 2\pi]$ is the noise strength.

We implement the backward updating method of the original Vicsek model to update the position $\boldsymbol{r}_i$ of particle $i$ according to

$$\boldsymbol{r}_i(t + \Delta t) = \boldsymbol{r}_i(t) + \boldsymbol{v}_i(t)\Delta t, \tag{17}$$

where $\boldsymbol{r}_i(t)$ represents the position vector of particle $i$, and $\boldsymbol{v}_i(t) = \left[v\cos\left(\theta_i(t)\right), v\sin\left(\theta_i(t)\right)\right]$ is the velocity vector of particle $i$ with constant speed $v$. Simulations are conducted in a square domain of size $L$ with the periodic boundary condition. The system is parameterized by the number of particles $N$, the particle density $\rho = N/L^2$, the neighborhood cutoff radius $R$, interaction weights $w_{ij}$, and the noise strength $\eta$. We set the constant speed to $v = 2$ and the time step to $\Delta t = 0.05$, consistent with the original Vicsek model, where results are robust for $0.003 < v\Delta t < 0.3$ [33]. To ensure sufficient statistics over initial conditions and over time, for each parameter set, we perform 65 independent simulation runs, each of 2000 time steps; the time-series data from these runs are concatenated, and the quantities of interest are computed. The above procedure is repeated independently five times to obtain reliable estimates of the mean and standard deviation.

In our simulations, the particles are divided into two equally sized sub-populations: influencers and followers. The index sets for the influencers and the followers are denoted by $\Omega_I$ and $\Omega_F$, respectively. We use $w_{I\to F}$ to represent the weights $w_{j\to i}$ for all $j \in \Omega_I$ and $i \in \Omega_F$. We fix $w_{I\to I} = w_{F\to I} = w_{F\to F} = 1$ and vary $w_{I\to F}$ and $\eta$ to induce the transitions among three characteristic phases: aligned, chiral, and disordered. Fig. 3(a) shows the three types of phase transitions: (1) when $w_{I\to F} \geq 0$, varying $\eta$ induces a transition between the aligned and the disordered phase, analogous to the transition in the standard Vicsek model; (2) when $w_{I\to F} < 0$, varying $\eta$ induces a transition between the chiral and the disordered phases; (3) when $\eta$ is small (low noise), varying $w_{I\to F}$ induces a transition between the aligned and the chiral phases. Fig. 3(b) shows the qualitative sketch of the phase diagram in the $\eta$–$w_{I\to F}$ plane.

We use three classical order parameters to characterize the three phase transitions, as shown in Fig. 4. The natural order parameter for the Vicsek model is the mean speed $v_a$ at each time step [33],

$$v_a(t) = \frac{1}{Nv}\left|\sum_i \boldsymbol{v}_i(t)\right|, \tag{18}$$

where $N$ is the number of particles. Denoting $T$ as the number of time steps, we then calculate the time average of the mean speed $\langle v_a\rangle_T$, the susceptibility $\chi$, and the Binder cumulant $G$:

$$\langle v_a\rangle_T = \frac{1}{T}\sum_t \big(v_a(t)\big), \tag{19}$$

$$\chi = L^2(\langle v_a^2\rangle_T - \langle v_a\rangle_T^2), \tag{20}$$

$$G = 1 - \frac{\langle v_a^4\rangle_T}{3\langle v_a^2\rangle_T^2}. \tag{21}$$

The susceptibility $\chi$ and Binder cumulant $G$ are the useful numerical estimation measures derived from the moments of the order parameter distribution used in the finite-size scaling approach [45] to help locate the transition points and characterize the type of phase transition. We emphasize that all three order parameters are derived from $v_a(t)$, which is based on the velocities/orientations of individual particles and does not contain explicitly the differences in velocities/orientations between pairs of particles.

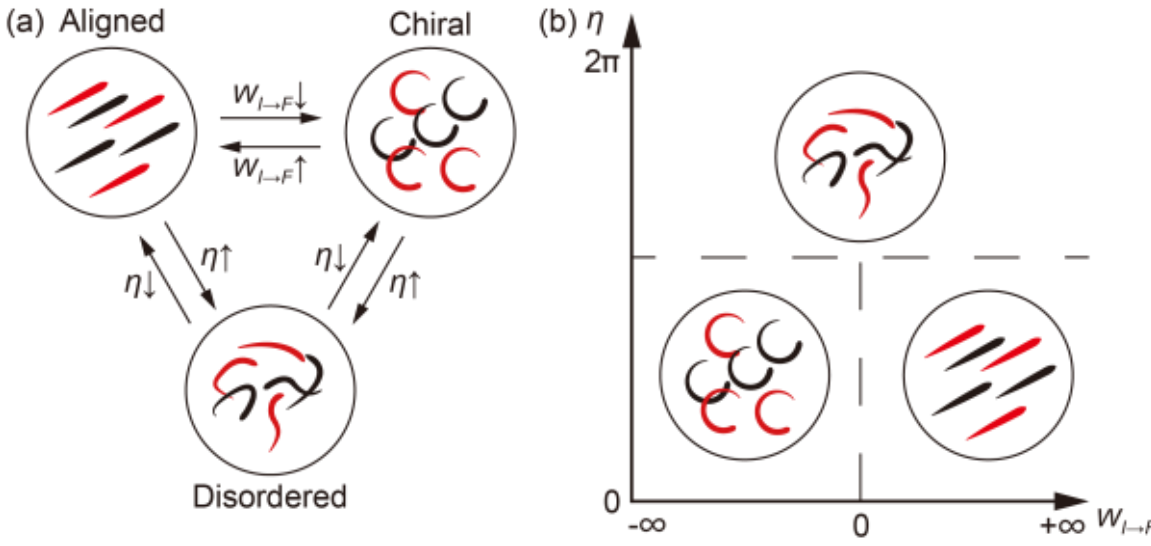

FIG. 3. Schematic of the phase transitions investigated in this study. (a) Relationship among the aligned, chiral, and disordered phases. (b) A qualitative sketch of the phase diagram in the $\eta - w_{I\to F}$ plane. The transitions between the aligned and chiral phases occur at small $\eta$ and are induced by changing signs of $w_{I\to F}$; the transitions between the aligned and disordered phases occur when $w_{I\to F} \geq 0$ and are induced by varying $\eta$; the transitions between the chiral and disordered phases occur when $w_{I\to F} < 0$ and are induced by varying $\eta$, too. See Movie S1 for representative videos of the transitions.

For the aligned-disordered transition [Figs. 4(a)-(c)], we set $w_{I\to F} = 1$ and $R = 5$ while varying $\eta$. With $w_{I\to F} = 1$, the interaction symmetry eliminates the distinction between influencers and followers. The chosen interaction radius $R = 5$ avoids clustering phenomena that occur at smaller radii so that the aligned phase emerges globally (other parameters are explored in Appendix B). Fig. 4(a) demonstrates that the modified model exhibits an order-disorder transition with increasing noise $\eta$, similar to the original Vicsek model [33]. The susceptibility peak in Fig. 4(b) indicates the critical noise is around $\eta = 1.2\pi$. The continuous evolutions of the order parameter $\langle v_a\rangle_T$ and the Binder cumulant in Figs. 4(a) and 4(c) are characteristics of a continuous-like phase transition for the system size considered. In the standard Vicsek model, a sharp drop of the Binder cumulant $G$ to negative values signals a first-order transition associated with large scale flocks and the emergence of high-density bands [45,46]. Such high-density bands are known to occur only when the system size is sufficiently large for a fixed interaction radius $R$ [45]. In our simulations, with $N = 400$, $L = 10$, $R = 5$, and a fixed particle density $\rho = 4$, $G$ remains positive across the entire noise range. The transition appears continuous within the accessible small-size regime, likely due to the finite-size effect. We provide a brief discussion and some preliminary results on finite-size scaling analysis in Appendix B. We also note that high-density bands do occur in our model for sufficiently large system sizes.

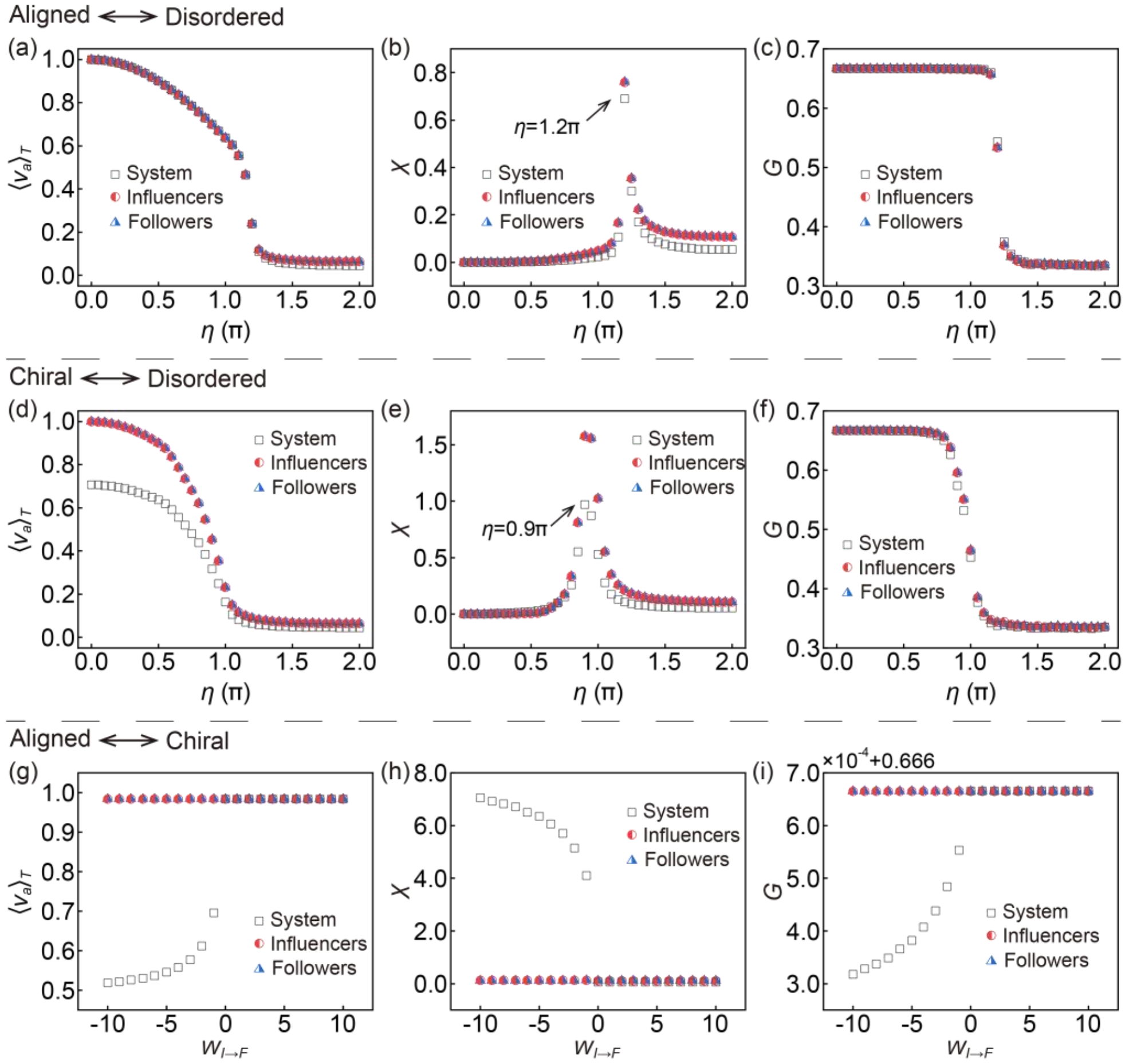

FIG. 4. Characterization of the three phase transitions using the classical order parameters, the time averages of the mean speed $\langle v_a \rangle_T$, the susceptibility $\chi$, and the Binder cumulant $G$. We calculate the order parameters for the entire population and for each sub-population. (a)–(c) The transition between the aligned and disordered phases with the $w_{I\to F} = 1$, $N = 400$, $\rho = 4$, and $R = 5$. (a) $\langle v_a \rangle_T$ versus $\eta$, (b) $\chi$ versus $\eta$, and (c) $G$ versus $\eta$. The peak of $\chi$ at $\eta = 1.2\pi$ indicates the transition point. The absence of a sharp drop in $G$ to negative values shows the behavior of a continuous-like transition at this finite system size. (d)–(f) The transition between the chiral and disordered phases with $w_{I\to F} = -1$, $N = 400$, $\rho = 4$, and $R = 5$. (d) $\langle v_a \rangle_T$ versus $\eta$, (e) $\chi$ versus $\eta$, and (f) $G$ versus $\eta$. The peak of $\chi$ at $\eta = 0.9\pi$ indicates the transition point. The absence of a sharp drop in $G$ to negative values shows the behavior of a continuous-like transition at this finite system size, too. (g)–(i) The transition between the aligned and chiral phases with $\eta = 0.2\pi$, $N = 400$, $\rho = 4$, and $R = 5$. (g) $\langle v_a \rangle_T$ versus $w_{I\to F}$. (h) $\mathcal{X}$ versus $w_{I\to F}$. (i) $G$ versus $w_{I\to F}$. The nature of this transition is an exceptional transition. The error bars are not shown as the sizes of the errors are smaller than the sizes of the symbols.

For the chiral-disordered transition [Figs. 4(d)-(e)], we set $w_{I\to F} = -1$ and $R = 5$ while varying $\eta$. A large radius $R$ enhances the many-body effect of the non-reciprocal interaction and leads to the emergence of a clear chiral phase, whereas a smaller value of $R$ causes the particles to aggregate in small local sub-populations and suppresses the formation of a global chiral phase. Since the collective splits into two sub-populations, we also plot the order parameter computed for each sub-population separately. Fig. 4(d) reveals that particles align within their own sub-population while maintaining a finite angular separation from the other sub-population. The global order decays continuously as noise increases. The peak in susceptibility [Fig. 4(e)] locates the critical point at $\eta = 0.9\pi$, and the continuous change in $G$ shows this as a continuous-like transition at this finite system size [Fig. 4(f)]. Comparing Figs. 4(a)-(c) and 4(d)-(f) show that the noise-driven transition exhibits the same continuous-like transition in both the aligned and chiral regimes for the system size considered.

For the aligned-chiral transition [Figs. 4(g)-(i)], we fix a low noise strength ($\eta = 0.2\pi$) to maintain an ordered state and set $R = 5$, sweeping $w_{I\to F}$ from $-10$ to $10$. Figs. 4(g)-(i) shows that the transition point occurs near $w_{I\to F} = 0$. This transition is an exceptional phase transition [47], as the sign of $w_{I\to F}$ determines whether the interactions from influencers to followers are reciprocal (positive, leading to alignment) or non-reciprocal (negative, leading to chiral motion).

We have so far characterized the phase transitions in our modified Vicsek model using the classical order parameters $\langle v_a \rangle_T$, $\chi$, and $G$. They are all based on instantaneous velocities/orientations of individual particles. However, the quantities that we are most interested in, namely, influence and information transfer, are inherently pairwise. Influence depends explicitly on the angular difference between two particles, and information transfer, according to our definitions in Sec. II, involves two particles, too. (Calculations of information transfer involving more than two particles are possible but not considered here.) This distinction not only motivated our quantifiable definition of influence, but also suggests that the study of influence and information transfer may reveal a richer understanding of the system's dynamics.

## IV. PAIRWISE INFLUENCE AND INFORMATION TRANSFER

We begin our analysis of influence and information transfer using a simple two-particle system [Fig. 5]. In this system, the interaction weight $w_{I\to F}$ from an influencer $I$ to a follower $F$ is varied from -100 to 100 and unless otherwise stated, all other weights are set to 1, i.e., $w_{I\to I} = w_{F\to I} = w_{F\to F} = 1$. The noise strength is varied from 0 to $2\pi$. In this section, we set $R = L$ so that the two particles are always interacting with each other.

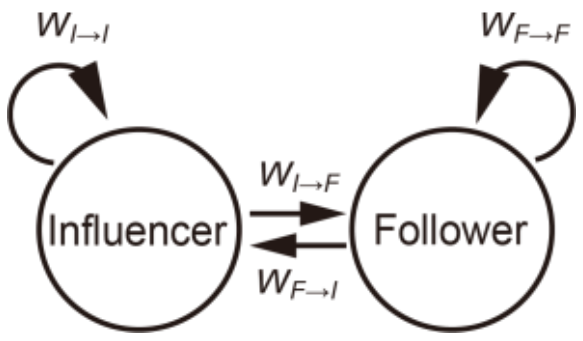

FIG. 5. The diagram of the two-particle system. Arrows indicate the directions of weights. Unless otherwise stated, in this section, we set $w_{I\to I} = w_{F\to I} = w_{F\to F} = 1$ and vary $w_{I\to F}$ and $\eta$ to tune their interaction. See Movie S2 for representative videos.

To quantify influence over time, we note that the direct averaging of the pairwise influences $\langle A_{I\to F} \rangle_T$ is close to zero. This cancellation, in the aligned and disordered phases, is due to the symmetry of the noise distribution, whereas in the chiral phase, it is due to the equal appearance of clockwise and anti-clockwise phases when combining independent simulations. Therefore, we calculate the time-averaged absolute influence, $\langle |A_{I\to F}| \rangle_T$,

$$\langle |A_{I\to F}| \rangle_T = \frac{1}{T} \sum_t |A_{I\to F}(t)|\,. \tag{22}$$

Simulation results of $\langle |A_{I\to F}| \rangle_T$, shown by the symbols in Figs. 6(a) and 6(b), indicate that the influence $\langle |A_{I\to F}| \rangle_T$ is a non-decreasing function of both the noise strength $\eta$ and the absolute value of weight $|w_{I\to F}|$. Fig. 6(b) shows that the dip at $w_{I\to F} = 0$ is consistent with the transition point of the aligned-chiral phases. It suggests that this phase transition can be seen clearly at the level of pairwise interaction, which suggests that this is qualitatively different from the other two transitions, i.e., the change from non-reciprocal to reciprocal interactions.

By setting the interaction radius of the particles equal to the system size ($R = L$), we simplify the subsequent analysis and are able to derive analytical expressions of influence (see Appendix C for details): When $w_{I\to F} \geq 0$,

$$\langle |A_{I\to F}| \rangle_T = \begin{cases} \dfrac{w_{I\to F}}{w_{I\to F} + w_{F\to F}} \times \dfrac{\eta}{3}, & \text{when } 0 \leq \eta \leq \pi, \\ \dfrac{w_{I\to F}}{w_{I\to F} + w_{F\to F}} \times \dfrac{-\eta^3 + 6\pi\eta(\eta - \pi) + 2\pi^3}{3\eta^2}, & \text{when } \pi < \eta \leq 2\pi. \end{cases} \tag{23}$$

Eq. (23) indicates that the slope at low noise is close to $1/3$ when $w_{I\to F}$ is large and the plateau value reached near $\eta \approx 2\pi$ is close to $\pi/2$ for large values of $w_{I\to F}$. When $w_{I\to F} < 0$, the value of $\langle |A_{I\to F}| \rangle_T$ is given by Eq. (24), where $m = \frac{|w_{I\to F}|}{1+|w_{I\to F}|}$ and $W = \frac{1+|w_{I\to F}|}{1+3|w_{I\to F}|}$. For $w_{I\to F} = -1$, we find $\langle |A_{I\to F}| \rangle_T = \pi/4$, a result that is notably independent of the noise strength $\eta$ as shown by the orange triangles in Fig. 6(a). For $\eta \to 2\pi$ and $|w_{I\to F}| \to \infty$, $\langle |A_{I\to F}(t)| \rangle_T \to \pi/2$.

$\langle |A_{I\to F}| \rangle_T$ as an increasing function of $|w_{I\to F}|$ for both $w_{I\to F} > 0$ and $w_{I\to F} < 0$ [Fig. 6(b)] can be explained as follows: a higher value of $|w_{I\to F}|$ means higher interaction strength giving higher $A_{I\to F}$ and hence higher $\langle |A_{I\to F}| \rangle_T$ as shown in Eq. (14). $\langle |A_{I\to F}| \rangle_T$ as a non-decreasing function of $\eta$ can be explained as follows: higher $\eta$ means that on average, a larger noise is added to the follower's present $\theta_F(t)$, as well as the influencer's present $\theta_I(t)$, according to Eq. (16). Consequently, within our model, this larger noise leads to a statistically larger difference on average between the influencer's future $\theta_I(t + \Delta t)$ and the follower's future $\theta_F(t + \Delta t)$, and hence

$$
\langle |A_{I\to F}(t)| \rangle_T = \begin{cases} mW\pi, & 0 \le \eta \le W\pi, \\ \dfrac{m}{3\eta^2}(-\pi^3 W^3 + 3\pi^2 W^2 \eta + \eta^3), & W\pi < \eta \le \pi - W\pi, \\ \dfrac{m\pi}{3\eta^2}\big(\pi^2(1 + W(-3 + (3 - 2W)W)) + 3\pi(-1 + 2W)\eta - 3(-1 + W)\eta^2\big), & \pi - W\pi < \eta \le W\pi + \pi, \\ \dfrac{m}{3\eta^2}(\pi^3(2 - (-6 + W)W^2) - 3\pi^2(2 + W^2)\eta + 6\pi\eta^2 - \eta^3), & W\pi + \pi \le \eta \le 2\pi - W\pi, \\ \dfrac{m\pi}{\eta^2}(\pi^2(-2 + 4W) + \pi(2 - 4W)\eta + W\eta^2), & 2\pi - W\pi < \eta \le 2\pi. \end{cases} \tag{24}
$$

produces a higher future influence $A_{I\to F}(t + \Delta t)$, according to Eq. (14). However, for the same $| w_{I\to F} |$ and small $\eta$, the case $w_{I\to F} < 0$ yields larger values because the chiral phase exhibits larger angular differences.

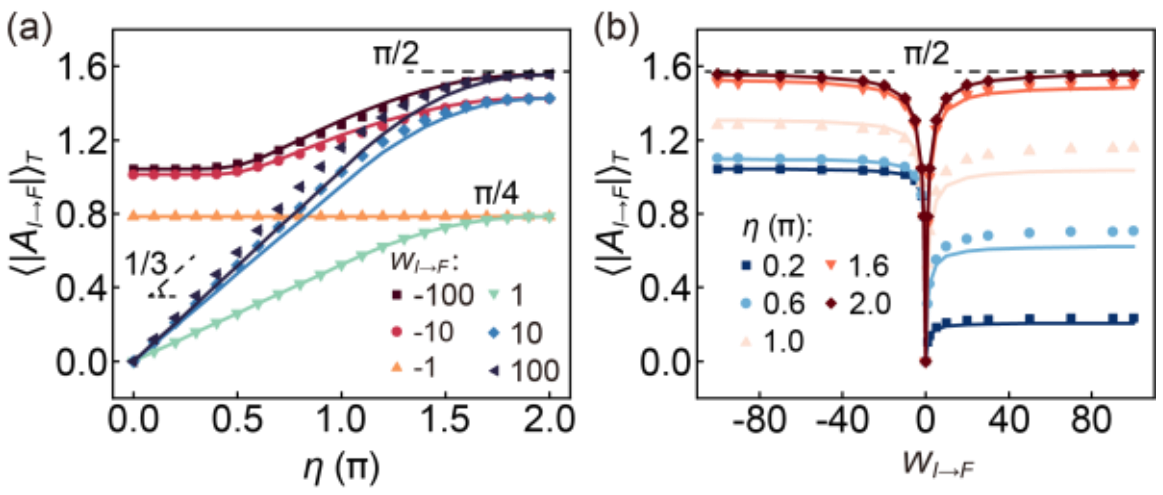

FIG. 6. Time-averaged absolute influence $\langle |A_{I\to F}| \rangle_T$ in the two-particle system. Symbols are simulation results; lines are analytical results calculated according to Eqs. (23) and (24). (a) $\langle |A_{I\to F}| \rangle_T$ versus $\eta$ for $w_{I\to F} = -100, -10, -1, 1, 10,\ 100$. For large $w_{I\to F}$, $1/3$ is the asymptotic slope in the low noise region, and $\pi/2$ is the asymptotic value as $\eta \to 2\pi$. $\langle |A_{I\to F}| \rangle_T$ is an increasing function of $\eta$, except in the case $w_{I\to F} = -1$, where it remains constant at $\pi/4$. (b) $\langle |A_{I\to F}| \rangle_T$ versus $w_{I\to F}$ for $\eta = 0.2\pi, 0.6\pi, 1.0\pi, 1.6\pi, 2.0\pi$. For $\eta = 2.0\pi$, $\pi/2$ is asymptotic value as $w_{I\to F} \to \pm\infty$. $\langle |A_{I\to F}| \rangle_T$ is an increasing function of $|w_{I\to F}|$ for both $w_{I\to F} > 0$ and $w_{I\to F} < 0$. However, for the same $| w_{I\to F} |$ and small $\eta$, the case $w_{I\to F} < 0$ yields larger values because the chiral phase exhibits larger angular differences.

To measure the information transfer from the influencer to the follower, we compute the TE from the influencer's present, $\theta_I(t)$, to the follower's future, $\theta_F(t + \Delta t)$, as shown in Figs. 7(a) and 7(b). Fig. 7(a) reveals two distinct trends: for $w_{I\to F} > 0$, TE initially increases and then decreases with increasing noise strength $\eta$; for $w_{F\to I} < 0$, it decreases with increasing $\eta$ for $\eta > 0$. The functional dependence of TE on $|w_{I\to F}|$ is similar to that of $\langle |A_{I\to F}| \rangle_T$ [Fig. 7(b)]: both quantities increase rapidly at small weights and reach plateau values at high weights. TE also increases monotonically with $|w_{I\to F}|$ for both positive and negative values of $w_{I\to F}$ and for the same $|w_{I\to F}|$ and small $\eta$, the case $w_{I\to F} < 0$ yields larger TE. Larger TE occurs because the chiral phase enriches the values of $\theta$ and their distribution, thereby enhancing information transfer. This dependence of TE on $w_{I\to F}$ aligns with intuition; the dependence of TE on $\eta$, however, is surprising. Given that $\langle |A_{I\to F}| \rangle_T$ increases monotonically with $\eta$, we had expected the similar monotonic increase for the TE. Instead, for $w_{I\to F} > 0$, TE peaks and then declines as $\eta$ increases and, for $w_{I\to F} < 0$, TE is decreasing for $\eta > 0$. What causes the behavior of TE shown in Fig. 7(a)?

The first clue is given by the total MI, $I\big(\theta_F(t), \theta_I(t); \theta_F(t + \Delta t)\big)$ [Fig. 7(c)]. The total MI is between the follower's future, $\theta_F(t + \Delta t)$, and both the follower's and the influencer's presents, $\theta_I(t)$ and $\theta_F(t)$. It decreases monotonously as $\eta$ increases. This monotonous decrease of the total MI shows that the noise suppresses the total information transfer between both the influencer's and the follower's presents and the follower's future. Indeed, we note that in Eq. (16), the follower's future $\theta_F(t + \Delta t)$ is a sum of three random variables: the follower's present $\theta_F(t)$, the present influence $A_{I\to F}(t)$, and the noise $\beta_F(t)$. The first two random variables are determined by the influencer's and the follower's present completely. Therefore, as the noise strength $\eta$ increases, the third term $\beta_F(t)$ becomes more dominant and suppresses the information transfer from the first two terms to the follower's future $\theta_F(t + \Delta t)$.

Another important clue is given by the normalized TE, which is the TE divided by the total MI [Fig. 7(d)]. The normalized TE increases as $\eta$ increases in the intermediate region of $\eta$ ($0 < \eta < 2\pi$). It suggests that increasing noise raises the relative importance of the influencer's present, as compared with the follower's present.

To gain further insights into the effects of noise, we turn off separately the noise of the influencer by setting $\eta_I = 0$ or the noise of the follower by setting $\eta_F = 0$ ,while keeping $w_{F\to I} = w_{I\to I} = w_{F\to F} = 1$ for all data shown in Fig. 8.

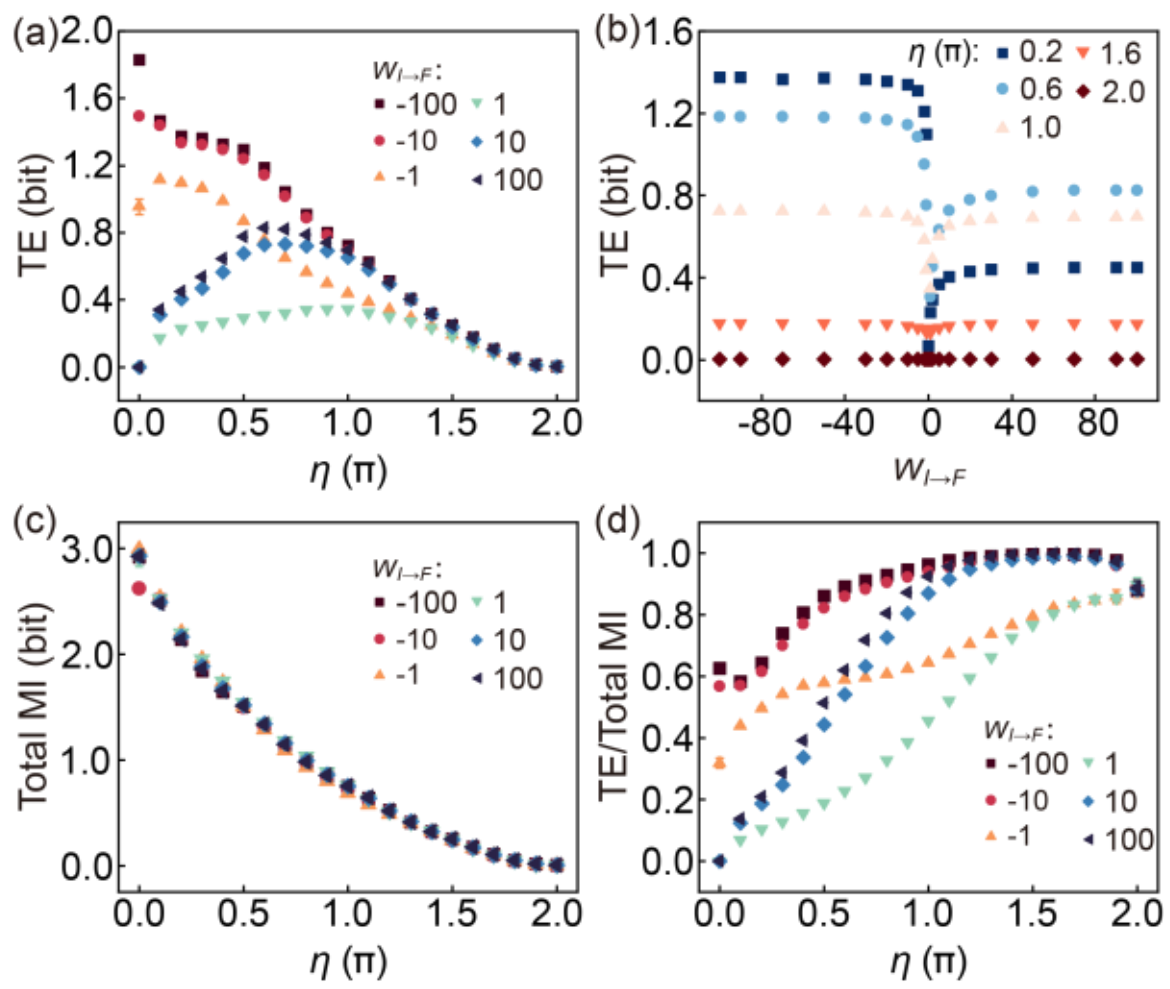

FIG. 7. Information transfer in the two-particle system. (a) TE from the influencer's present to the follower's future as a function of $\eta$ for $w_{I\to F} = -100, -10, -1, 1, 10, 100$. For $w_{I\to F} > 0$, the TE exhibits a non-monotonic dependence on $\eta$, which hints at the dual effects of noise. For $w_{I\to F} < 0$, TE decreases with increasing $\eta$ for $\eta > 0$. (b) TE versus $w_{I\to F}$ for $\eta = 0.2\pi, 0.6\pi, 1.0\pi, 1.6\pi, 2.0\pi$. TE is an increasing function of $|w_{I\to F}|$ for both $w_{I\to F} > 0$ and $w_{I\to F} < 0$. However, for the same $|w_{I\to F}|$ and small $\eta$, the case $w_{I\to F} < 0$ yields larger TE. This is because the chiral phase enriches the values of $\theta$ and their distribution, thereby enhancing information transfer. (c) Total MI between both the influencer's and the follower's presents and the follower's future as a function of $\eta$. The total MI decreases with increasing $\eta$. (d) Normalized TE (TE divided by total MI) as a function of $\eta$. The normalized TE increases as $\eta$ increases in the intermediate region of $\eta$ ($0 < \eta < 2\pi$).

In Fig. 8(a), we study the case of $w_{I\to F} = 1$. When $\eta_I \neq 0$ and $\eta_F = 0$, according to Eq. (16), the follower's future $\theta_F(t + \Delta t)$ becomes the sum of only two random variables, the follower's present $\theta_F(t)$ and the present influence $A_{I\to F}(t)$. Increasing the noise strength of the influencer $\eta_I$ widens the range of influencer's orientations, thus increasing the amount of information generated by the influencer at each step and, consequently, increasing the information transfer from the influencer to the follower, as shown by the grey symbols in Fig. 8(a). When $\eta_I = 0$ and $\eta_F \neq 0$, the value of TE is smaller overall than in the case of $\eta_F = 0$ and exhibits a non-monotonic behavior, roughly first increasing and then decreasing. When $\eta_F = \eta_I = 0$, both influencer and follower move linearly, and $\theta_I = \theta_F$; hence, no information needs from the influencer to be transferred to the follower, and TE is 0. As $\eta_F$ increases, the follower's orientation begins to fluctuate, and the influencer attempts to realign it—a process that transfers a small amount of information, causing TE to rise slightly. At larger $\eta_F$, the follower's dynamics become dominated by noise, which suppresses the information received from the influencer, and TE declines.

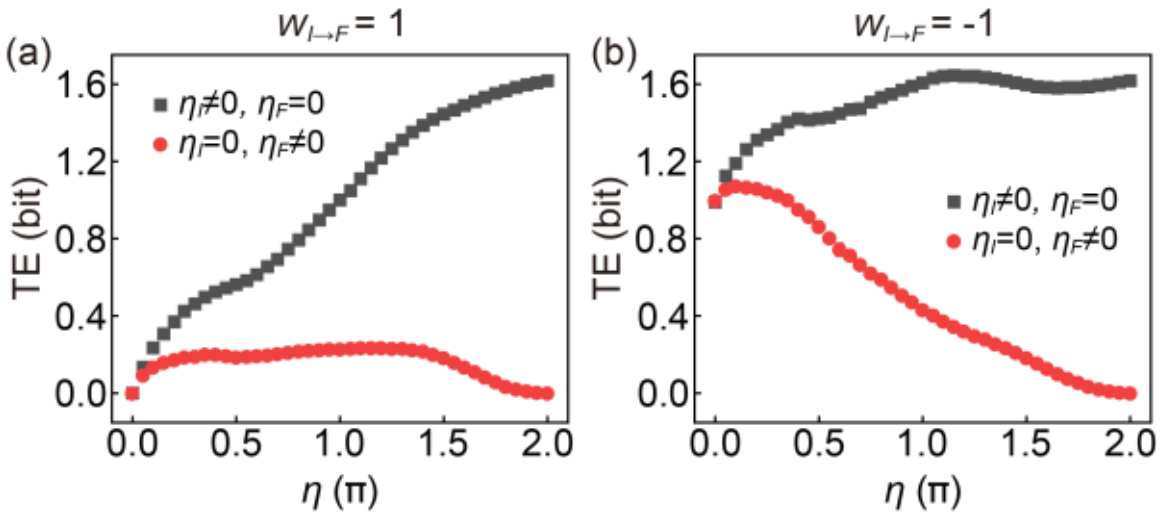

FIG. 8. Investigation of the dual effects of noise on information transfer. We turn off the influencer's noise $\eta_I$ or the follower's noise $\eta_F$ separately and keep $w_{F\to I} = w_{I\to I} = w_{F\to F} = 1$ . (a) TE from the influencer's present to the follower's future as a function of $\eta_I$ or $\eta_F$ for the case $w_{I\to F} = 1$. (b) TE versus $\eta_I$ or $\eta_F$ for the case $w_{I\to F} = -1$. The comparison between the cases with only influencer's noise $\eta_I$ and the cases with only follower's noise $\eta_F$ suggests that the overall effect of $\eta_I$ is enhancing information transfer, whereas the overall effect of $\eta_F$ is suppressing information transfer.

In Fig. 8(b), we study the case of $w_{I\to F} = -1$. When $\eta_I \neq 0$ and $\eta_F = 0$, increasing $\eta_I$ initially increases TE first, and then TE saturates after $\eta_I = \pi$. In the chiral phase, without noise, the angular difference between the influence and the follower is a constant. Adding a small amount of noise introduces fluctuations in the influencer's orientation, enriching its behavioral variability while preserving the chiral phase. These fluctuations create variations in the angular difference between influencer and follower, thereby increasing information transfer. Conversely, excessive noise destabilizes the chiral phase, driving the system into disorder and suppressing information transfer. The competition between enriching orientational differences (which can increase information transfer) and disrupting the chiral phase (which suppresses it) leads to the saturation for large $\eta_I$ . When $\eta_I = 0$ and $\eta_F \neq 0$, the value of TE decreases nearly monotonically with $\eta_F$, as the follower's own noise progressively undermines the ordered chiral dynamics.

In summary, increasing the noise of the influencer ($\eta_I$) increases the TE from the influencer to follower, whereas increasing noise on the follower ($\eta_F$) decreases TE—particularly apparent at high noise levels where TE reduces to zero. Thus, combining the observation in Figs. 7(c) and 7(d), we conclude that the noise on the influencer can increase the information transfer from the influencer to the follower, whereas the noise on the follower will decrease the information received by the follower. It has also been reported that information transfer from a leader to a follower in the Vicsek model increases with noise at low noise strengths and decreases at high noise strengths [19], which is consistent with the dual effects we observe in our modified Vicsek model. This enriching effect of noise is counterintuitive and reminiscent of the role of noise in Maxwell's demon and Szilard's engine [48–50]. Indeed, it is not unusual for noise to play opposing roles depending on its magnitude. It is well known that strong noise can disrupt order. Conversely, weak noise has been reported to drive a system out of an inefficient deterministic regime and into a more ordered one, thereby facilitating order [44]. Moreover, a phenomenon known as "order by disorder" describes the counterintuitive process in which fluctuations, rather than destroying order, lift the massive ground-state degeneracy of a frustrated magnet and induce a net ordered phase [51–54].

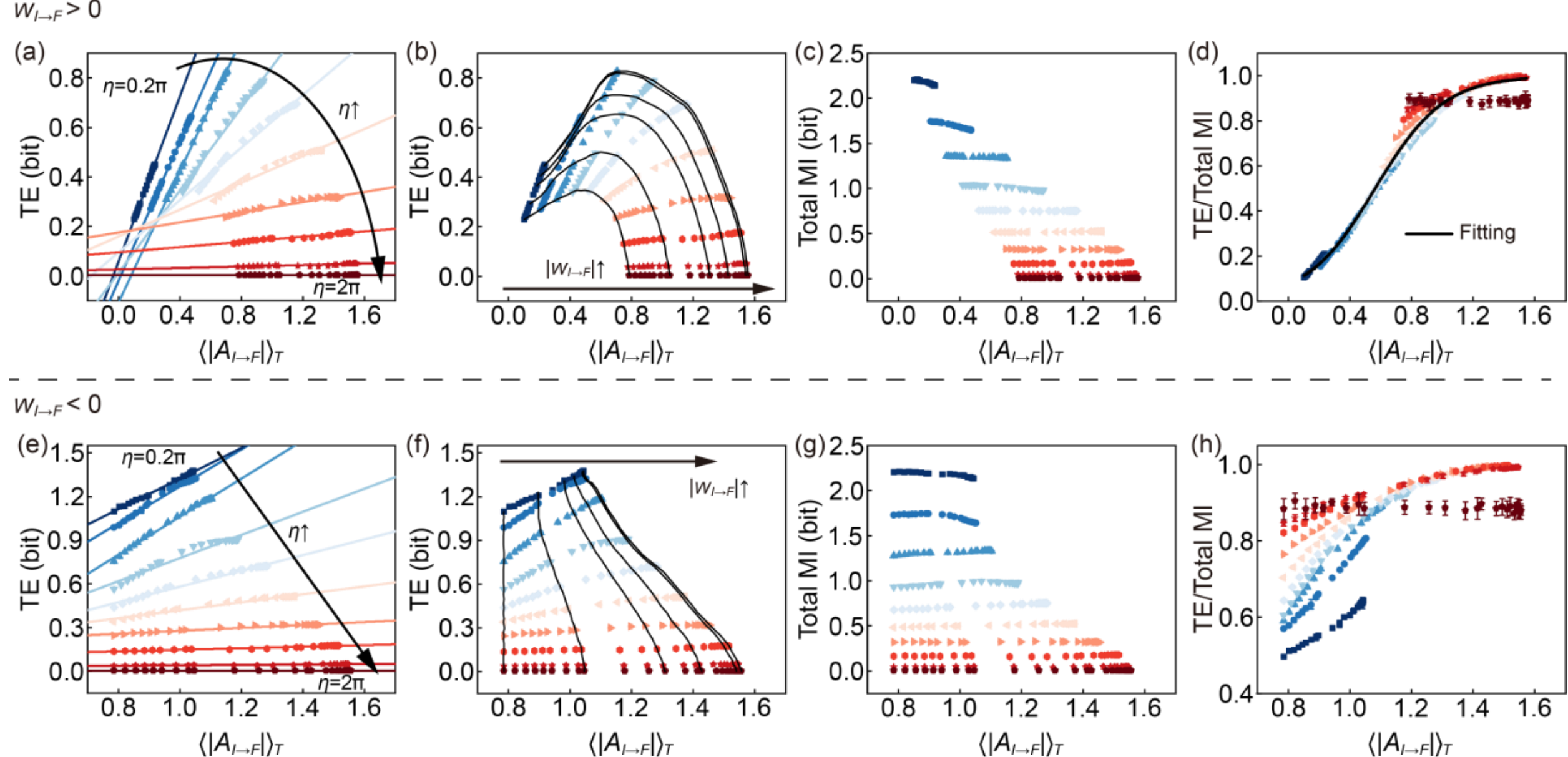

FIG. 9. Information transfer as a function of influence $\langle|A_{I\to F}|\rangle_T$ in the two-particle system. Each node is color-coded according to its noise strength, with different colors representing different values of $\eta$, ranging from $0.2\pi$ to $2.0\pi$. Symbols of the same color represent different values of $w_{I\to F}$. (a)–(d) $w_{I\to F} > 0$. (a) TE versus $\langle|A_{I\to F}|\rangle_T$. Lines are linear fittings. For each fixed $\eta$, TE exhibits a quasi-linear relation with $\langle|A_{I\to F}|\rangle_T$. (b) TE versus $\langle|A_{I\to F}|\rangle_T$. Each curve represents how TE varies with $\langle|A_{I\to F}|\rangle_T$ at the same $w_{I\to F}$ ($w_{I\to F} = 1, 2, 5, 10, 50,$ and $100$). $\langle|A_{I\to F}|\rangle_T$ display nonlinear relations with the pairwise TE because of the dual effects of noise. (c) Total MI versus $\langle|A_{I\to F}|\rangle_T$. Total MI decreases with increasing $\eta$. At each fixed $\eta$, total MI changes slightly. (d) Normalized TE versus $\langle|A_{I\to F}|\rangle_T$. The curve is the fitting of a Boltzmann sigmoid function for data excluding the case $\eta = 2\pi$. (e)–(h) $w_{I\to F} < 0$. (e) TE versus $\langle|A_{I\to F}|\rangle_T$. Lines are linear fittings. For each fixed $\eta$, TE shows a quasi-linear relation with $\langle|A_{I\to F}|\rangle_T$. (f) TE versus $\langle|A_{I\to F}|\rangle_T$. Each curve represents how TE varies with $\langle|A_{I\to F}|\rangle_T$ at the same $w_{I\to F}$ ($w_{I\to F} = -1, -2, -5, -10, -50,$ and $-100$). $\langle|A_{I\to F}|\rangle_T$ display nonlinear relations with the pairwise TE. (g) Total MI versus $\langle|A_{I\to F}|\rangle_T$. Total MI decreases with increasing $\eta$. At each fixed $\eta$, total MI changes slightly. (h) Normalized TE versus $\langle|A_{I\to F}|\rangle_T$. For each $\eta$, normalized TE increases with $\eta$ except in the case $\eta = 2\pi$, where it remains relatively constant.

Finally, we analyze the direct relations between information transfer and influence [Fig. 9]. Figs. 9(a)-(d) correspond to the case $w_{I\to F} > 0$. Fig. 9(a) shows that at a fixed noise strength $\eta$, as $w_{I\to F}$ increases from 1 to 100, the TE and the time-averaged absolute influence $\langle|A_{I\to F}|\rangle_T$ exhibit a quasi-linear relation. This quasi-linear relationship indicates that at a fixed $\eta$, higher influence means more information transferred. Fig. 9(b) shows that, for a fixed weight, TE first increases and then decreases with increasing influence, mirroring its non-monotonic dependence on $\eta$ in Fig. 7(a). This non-linear relation is due to the dual effects of noise. Fig. 9(c) reveals that at a fixed $\eta$, the total MI decreases a little or stays constant as influence increases, but at a fixed weight, total MI decreases as influence increases. Fig. 9(d) shows that the normalized TE roughly increases as influence $\langle|A_{I\to F}|\rangle_T$ increases and roughly collapse onto a single sigmoid curve, except for the case when the noise strength $\eta = 2\pi$. If we remove the data points for $\eta = 2\pi$, the rest of the data points can be fitted with a Boltzmann sigmoid function [Fig. 9(d)]:

$$\text{TE/Total MI} = \frac{\exp(\langle |A_{I\to F}| \rangle_T / k)}{\exp(\langle |A_{I\to F}| \rangle_T / k) + \exp(a/k)}, \quad (25)$$

and the fitting parameters $a = 0.564$ and $k = 0.223$. The parameter $a$ can be regarded as the self-influence of the follower, and $k$ can be regarded as a characteristic influence scale of the system. Thus, the two exponential terms, $\exp(\langle |A_{I\to F}| \rangle_T / k)$ and $\exp(a/k)$, represent the proportions of information transfer coming from the influencer's present and from the follower's present, respectively. As influence increases, the relative importance of the influencer increases as compared with the follower's self-influence.

Figs. 9(e)-(h) present the case when $w_{I\to F} < 0$. Fig. 9(e) shows that at a fixed $\eta$, as $w_{I\to F}$ varies from -1 to -100, the quasi-linear relations between TE and $\langle |A_{I\to F}| \rangle_T$ still exist. Thus, for a fixed $\eta$, higher influence means more information transferred. For a fixed $w_{I\to F}$, Fig. 9(f) shows that the increasing of $\eta$ decreases TE while increasing $\langle |A_{I\to F}| \rangle_T$, consistent with the trends in Figs. 6(a) and 7(a). Fig. 9(g) reveals that at a fixed noise strength $\eta$, the total MI decreases a little or stays constant as influence increases, whereas at a fixed weight, total MI decreases as influence increases. Fig. 9(h) indicates that the relationship between influence and TE no longer follows the Boltzmann sigmoid form. This breakdown stems from the information-enriching effect of the chiral phase itself, which arises from non-reciprocal interactions.

We summarize the key points to conclude this section on the two-particle system. First, our modified Vicsek model allows direct calculations of influence $\langle |A_{I\to F}| \rangle_T$. Second, our analysis reveals the dual effects of noise on information transfer for cases of $w_{I\to F} > 0$: increasing the influencer's noise increases the information transfer, whereas increasing the follower's noise suppresses the information transfer. Most significantly, we find quasi-linear relations between the transfer entropy (TE) and the influence $\langle |A_{I\to F}| \rangle_T$ at fixed $\eta$, as shown in Figs. 9(a) and 9(e), which provides validation for the effectiveness of TE as a measure of directed information flow at fixed noise strength in our system.

## V. COLLECTIVE INFLUENCE AND INFORMATION TRANSFER

After analyzing the pairwise influence and information transfer in the previous section, we analyze the collective influence and information transfer across three phase transitions. The simulations of the collective are the same as the data used in Fig. 4.

Given that particles are divided into two sub-populations—influencers and followers—under non-reciprocal interactions, and notice that the collective phase (especially the chiral phase) is mainly affected by the interaction between two sub-populations, we quantify the net influence exerted by the influencer sub-population on the follower sub-population. Starting from the pairwise influence $A_{j\to i}(t)$, we construct suitable averages of pairwise influence using four operations: the neighbor-average operation over influencer-neighbors within the cutoff radius $\langle \cdot \rangle_{R_I}$ where the subscript $R_I$ indicates the number of influencers who are neighbors of particle $i$; the number-average operation over followers $\langle \cdot \rangle_{N_F}$; the time-average operation $\langle \cdot \rangle_T$; and the norm operation $|\cdot|$. The permutations of the operations give 24 possible types of averages, but because the order of averaging operations can be switched and because the effects of the time-averaging and number-averaging are similar, there are only 4 significantly different averages. Using $N_F$ to represent the number of followers, we list the four significant averages below:

(1) $\langle |A| \rangle_{R_I,N_F,T}$: absolute pairwise influences averaged over influencer-neighbors, number of followers, and time:

$$\langle |A| \rangle_{R_I,N_F,T} = \frac{1}{T}\sum_{t=1} \frac{1}{N_F} \sum_{i\in\Omega_F} \sum_{j\in\Omega_I} |A_{j\to i}(t)|, \quad (26)$$

(2) $\langle |\langle A \rangle_{R_I}| \rangle_{N_F,T}$: absolute influencer-neighbor-averaged pairwise influences averaged over the number of followers and time:

$$\langle |\langle A \rangle_{R_I}| \rangle_{N_F,T} = \frac{1}{T}\sum_{t=1} \frac{1}{N_F} \sum_{i\in\Omega_F} \left| \sum_{j\in\Omega_I} A_{j\to i}(t) \right|, \quad (27)$$

(3) $\langle |\langle A \rangle_T| \rangle_{R_I N_F}$: absolute time-averaged influences averaged over neighbor influencers and the number of followers:

$$\langle |\langle A \rangle_T| \rangle_{R_I N_F} = \frac{1}{N_F} \sum_{i\in\Omega_F} \sum_{j\in\Omega_I} \left| \frac{1}{T}\sum_{t=1} A_{j\to i}(t) \right|, \quad (28)$$

(4) $|\langle A \rangle_{T,R_I,N_F}|$: absolute time-averaged, follower-number-averaged, and influencer-neighbor-averaged influence:

$$|\langle A \rangle_{T,R_I,N_F}| = \left| \frac{1}{N_F} \sum_{i\in\Omega_F} \sum_{j\in\Omega_I} \frac{1}{T}\sum_{t=1} A_{j\to i}(t) \right|. \quad (29)$$

The difference among the four unique ones is the position at which the absolute value is taken in the sequence of operations. $\langle |A| \rangle_{R_I,N_F,T}$ and $\langle |\langle A \rangle_{R_I}| \rangle_{N_F,T}$ are order-of-magnitude larger than $\langle |\langle A \rangle_T| \rangle_{R_I N_F}$ and $|\langle A \rangle_{T,R_I,N_F}|$ because averaging time steps zeroes out the noise fluctuations and eliminates the net influence over time due to the equal appearance of clockwise and anti-clockwise chiral phases. In addition, $\langle |A| \rangle_{R_I,N_F,T} \geq \langle |\langle A \rangle_{R_I}| \rangle_{N_F,T}$ and $\langle |\langle A \rangle_T| \rangle_{R_I N_F} \geq |\langle A \rangle_{R_I,N_F,T}|$ because $\sum |a| \geq |\sum a|$ for any real number $a$, so we have

$$\langle |A| \rangle_{R_I,N_F,T} \geq \langle |\langle A \rangle_{R_I}| \rangle_{N_F,T} \geq \langle |\langle A \rangle_T| \rangle_{R_I N_F} \geq |\langle A \rangle_{T,R_I,N_F}|, \quad (30)$$

The behaviors of influence-based order parameters in different phase transitions are shown in Fig. 10, and the specific form given in Eq. (30) is validated in Figs. 10(a), 10(c), and 10(f). The $\langle |\langle A \rangle_{R_I}| \rangle_{N_F,T}$ is the most interesting among those influence-based order parameters. For the aligned-disordered transition, this quantity exhibits an inverted V-shape near the critical point, resembling the curve of specific heat of a Weiss ferromagnet model as a function of temperature [55] and allowing for unambiguous identification of the transition. It also displays clear signatures across the chiral-disordered and aligned-chiral transitions. Moreover, averaging the influence over neighbors first, instead of applying the other 3 operations for the pairwise influence, is a more natural choice, as $\langle A \rangle_{R_I}$ aggregates all influences from influencer-neighbors onto a given follower, thereby capturing the collective effect. In contrast, the remaining operations focus on preserving pairwise details rather than this aggregate effect. Thus, we select $\langle |\langle A \rangle_{R_I}| \rangle_{N_F,T}$ as the primary focus of our discussion. Henceforth, unless specified otherwise, the term "influence" refers to this particular measure and its fluctuation in the study of collective behavior. Furthermore, we want to point out that $\langle |A| \rangle_{R_I,N_F,T}$ is more suitable to emphasize the pairwise influence in the collective. $\langle |A| \rangle_{R_I,N_F,T}$ in Fig. 10(a) and Fig. 10(c) is similar to the curve for $w_{I \to F} = 1$ and $w_{I \to F} = -1$ in Fig. 6(a), respectively. $\langle |\langle A \rangle_T| \rangle_{R_I N_F}$ is better suited to studies focused on the time evolution of pairwise influence and $|\langle A \rangle_{T,R_I,N_F}|$ is more appropriate when both neighbor averaging and temporal dynamics are relevant, as it applies the absolute value only after all other operations.

For the aligned-disordered transition, besides the inverted V-shaped, Fig. 10(b) shows that for $\eta \leq \pi$, the influence increases linearly with $\eta$, and saturates to an asymptotic value as $\eta \to 2\pi$. We summarize the analytical evaluation of this linear regime and asymptotic limit here (detailed in Appendix D). In the ordered state ($\eta \leq \pi$), we assume that the influences on one particle by all neighbors align the particle in the direction $\theta_{order}$ at each time step: $\theta_{order} = \theta_i(t) + A_i(t)$. Using this assumption, we derive the linear relation $\langle|\langle A\rangle_{R_I}|\rangle_{N_F,T} = \eta/8$, plotted as the line in Fig. 10(b), which agrees well with the simulations. The asymptotic value of $\langle|\langle A\rangle_{R_I}|\rangle_{N_F,T}$ as $\eta \to 2\pi$ corresponds to the expectation value of $|\langle A\rangle_R|$ under complete disorder. By the central limit theorem, $\langle A\rangle_{R_I}$ follows a normal distribution with zero mean and variance $\pi^2/(3n_I)$, where $\pi^2/3$ is the variance of a uniform distribution over $[-\pi,\pi]$, and $n = \pi R^2 \rho_I$ is the average number of neighbor influencers where $\rho_I$ is the density of influencers. Hence, $\langle|\langle A\rangle_{R_I}|\rangle_{N_F,T} = \sqrt{2/(3R^2\rho_I)} \propto 1/R$. In other words, for a fixed particle density $\rho$, the asymptotic value is inversely proportional to the interaction cutoff radius $R$. As the orientations of particles are dominated by the noises when $\eta \approx 2\pi$, increasing $R$ increases the number of influencer neighbors for each follower, driving the average influence from influencers toward zero in a statistical sense. This indicates that it becomes less possible for a follower to extract a more preferred orientation from the influencer-neighbors than its original orientation. The decay of influence with increasing $R$ scales as $1/R$, reflecting the rate of the influence decay, which is related to the rate of increasing the number of influencer-neighbors and governed by the central limit theory. When $R = 5$, $\rho = 4$, and $\rho_I = 2$, $\langle|\langle A\rangle_{R_I}|\rangle_{N_F,T} = 0.058$ which is consistence with the simulation results.

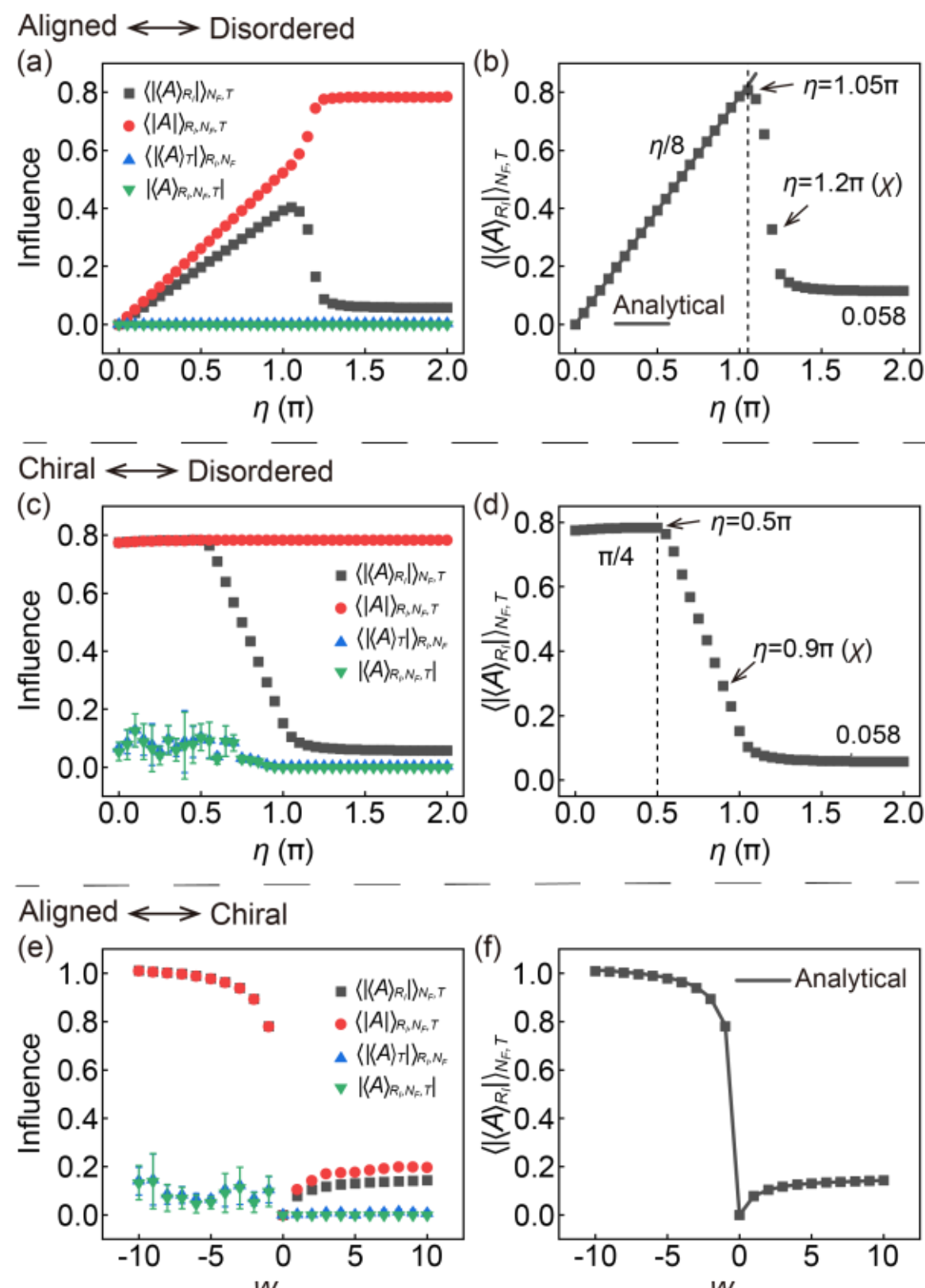

FIG. 10. Characterization of the phase transitions with influence-based order parameters. The simulations are the same as the ones in Fig. 4. (a)-(b) The transition between the aligned and disordered phases. (a) $\langle |A| \rangle_{R_I,N_F,T}$, $\langle |\langle A \rangle_{R_I}| \rangle_{N_F,T}$, $\langle |\langle A \rangle_T| \rangle_{R_I N_F}$, $|\langle A \rangle_{T,R_I,N_F}|$ as a function of $\eta$. The inverted V-shape of $\langle |\langle A \rangle_{R_I}| \rangle_{N_F,T}$ allows easy identification of the transition point. (b) $\langle |\langle A \rangle_{R_I}| \rangle_{N_F,T}$ versus $\eta$. The solid line is the analytical result for $\eta \leq \pi$; the peak position of $\langle |\langle A \rangle_{R_I}| \rangle_{N_F,T}$ is at $\eta = 1.05\pi$; the peak position of $\chi$ at $\eta = 1.2\pi$ is also labeled; the asymptotic value for $\eta \to 2\pi$ agrees with the analytical result of 0.058. (c)-(d) The transition between chiral and disordered phases. (c) The four influence-based order parameters versus $\eta$. The values of $\langle |\langle A \rangle_{R_I}| \rangle_{N_F,T}$ exhibit a clear transition. (d) $\langle |\langle A \rangle_{R_I}| \rangle_{N_F,T}$ versus $\eta$. For $\eta \leq 0.5\pi$, $\langle |\langle A \rangle_{R_I}| \rangle_{N_F,T}$ agrees with the analytical result of $\pi/4$. The peak position of $\chi$ at $\eta = 0.9\pi$ is also labeled. The asymptotic value for $\eta \to 2\pi$ agrees with the analytical result of 0.058. (e)-(f) The transition between aligned and chiral phases. (e) The four different influence-based order parameters versus $\eta$. (f) $\langle |\langle A \rangle_{R_I}| \rangle_{N_F,T}$ versus $\eta$. The solid line is the analytical result. The transition point at $w_{I \to F} = 0$ agrees with the transition point identified by susceptibility $\chi$.

For the chiral-disordered transition, we can analytically derive that when the noise strengths $\eta \leq \pi - \frac{|w_{I\to F}|}{1+3|w_{I\to F}|}\pi$, the influence is $\langle|\langle A\rangle_{R_I}|\rangle_{N_F,T} = \frac{|w_{I\to F}|}{1+3|w_{I\to F}|}\pi$. For the special case $w_{I\to F} = -1$, this reduces to $\eta \leq \pi/2$ and $\langle|\langle A\rangle_{R_I}|\rangle_{N_F,T} = \pi/4$. This derivation assumes the noise is sufficiently weak that it cannot disrupt the influence used to maintain the chiral phase. Thus, $\eta = \pi - \frac{|w_{I\to F}|}{1+3|w_{I\to F}|}\pi$ can be viewed as the threshold point beyond which the chiral phase starts to become chaotic. Fig. 10(d) is consistent with the analytical result, which identifies the transition point at $\eta = \pi/2$. Moreover, for $\eta \leq \pi/2$, $\langle|\langle A\rangle_{R_I}|\rangle_{N_F,T}$ increases slightly with $\eta$ and remains close to $\pi/4$.

It is noteworthy that the transition points identified by the influence measure in Figs. 10(b) and 10(d) differ from the critical points indicated by the susceptibility peaks in Figs. 4(b) and 4(e). This difference likely arises because the two metrics probe different stages of the transition process. The influence, as derived analytically, is sensitive to the initial loss of order—it effectively marks the onset of the transition where coherent motion begins to break down. In contrast, the susceptibility, which measures the variance of the order parameter, peaks at the midpoint of the transition, where fluctuations are maximal. Thus, while related, these quantities highlight distinct features of the phase transition dynamics. Fig. 10(d) also shows that the influence saturates to an asymptotic value as $\eta \to 2\pi$. Following the same reason at the aligned-disordered case, this asymptotic limit is $\langle|\langle A\rangle_{R_I}|\rangle_{N_F,T} = \frac{|w_{I\to F}|}{1+|w_{I\to F}|}\sqrt{\frac{2}{3R^2\rho_I}}$. For the parameters $w_{I\to F} = -1$, $R = 5$, and $\rho_I = \rho/2 = 4$, this evaluates to approximately 0.058, consistent with the numerical results.

For the aligned-chiral transition, $\langle|\langle A\rangle_{R_I}|\rangle_{N_F,T}$ clearly identifies the transition point at $w_{I\to F} = 0$. As the transition occurs at low noise, the influence can be obtained analytically throughout the process. The larger value of influence in chiral phase ($w_{I\to F} \leq 0$) compared to the aligned phase ($w_{I\to F} > 0$) arises from a fundamental difference in the interaction dynamics. In the chiral phase, the two sub-populations maintain a finite angular separation, resulting in continuous mutual influence. In contrast, in the aligned phase, the sub-populations are co-directional; once alignment is achieved, the primary role of influence is to counter the disordering effects of noise, which requires less sustained directional forcing. In summary, the source of information content is noise in the aligned phase, as revealed by our discussion in the previous section, whereas the source of information content in the chiral phase is the cycling motion of the particles themselves.

Next, we investigate collective information transfer. A recent report uses global transfer entropy to perform a similar analysis, but their analysis did not reveal any direct link between information transfer and the collective states [56]. Our results so far have shown that influence can indicate phase transitions [Fig. 10] and that information transfer and influence are quantitatively related [Fig. 9], so we ask the question: Can we infer phase transitions from information transfer? The answer is yes, and we provide our analysis below.

To compute the information received by a follower from its influencer-neighbors, we devise the following method to treat all influencer-neighbors as one entity. First, we collect triplets of angular orientations $\left(\theta_j(t), \theta_i(t), \theta_i(t+\Delta t)\right)$ for all neighboring pairs $(i,j)$ over the entire duration of a simulation; then we consider $\theta_j(t)'s$ of all the influencer-neighbors of follower $i$ as a single variable $\theta_{nbs}(t)$ and obtain the aggregated set of triplets $\left(\theta_{nbs}(t), \theta_i(t), \theta_i(t+\Delta t)\right)$ for each follower $i$; next, we construct the joint probability distribution $p\big(\theta_{nbs}(t), \theta_i(t), \theta_i(t+\Delta t)\big)$ and compute TE; then, we repeat the above calculation for 100 particles for enough statistics [Appendix E] and compute average TE; finally, we compute the average and standard deviations of average TEs from five independent simulation runs. By aggregating the orientations of all neighboring influencers' $\theta_j(t)$ into a single variable $\theta_{nbs}(t)$, we are effectively performing a reduction of complexity similar in spirit to the mean-field approximation in the Curie-Weiss theory on the Ising model. Our approach assumes that, for the purpose of quantifying information flow into one follower, the aggregated collective state of its influencer-neighborhood is the most relevant degree of freedom, removing irrelevant specific identities of individual influencer-neighbors. This simplification is both physically motivated and methodologically necessary to make the information-theoretic analysis of a many-body system computationally feasible. The process of constructing $\theta_{nbs}$ is also consistent with $\langle|\langle A\rangle_{R_I}|\rangle_{N_F,T}$, i.e., addressing neighbor's behavior first.

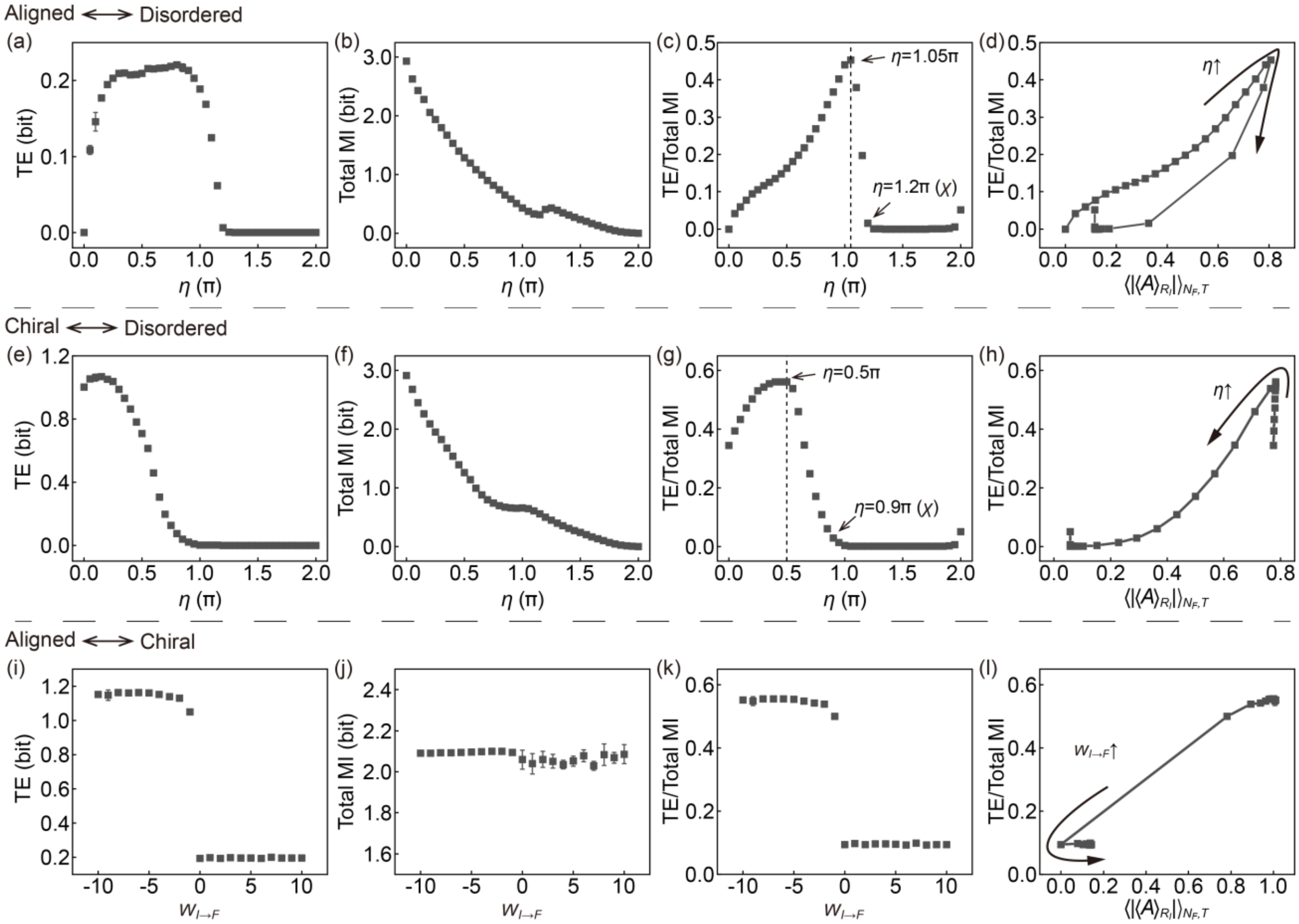

FIG. 11. Characterization of the phase transitions with information transfer. The simulations are the same as the ones in Fig. 4 and Fig. 10. The information transfer is measured from the followers' neighbors who are influencers (influencer-neighbors) to the followers. (a)–(d) Transition between the aligned and disordered phases. (a) TE versus $\eta$. (b) Total MI versus $\eta$. (c) Normalized TE (TE/Total MI) versus $\eta$. The peak position is at $\eta = 1.05\pi$, which agrees with the transition point identified by influence $\langle|\langle A\rangle_{R_I}|\rangle_{N_F,T}$ but differs from the transition point identified by susceptibility $\chi$ at $\eta = 1.2\pi$. (d) Normalized TE (TE/Total MI) versus $\langle|\langle A\rangle_{R_I}|\rangle_{N_F,T}$. (e)–(h) Transition between the chiral and disordered phases. (e) TE versus $\eta$. (f) Total MI versus $\eta$. (g) Normalized TE versus $\eta$. The peak position is at $\eta = 0.5\pi$, which agrees with the transition point identified by influence $\langle|\langle A\rangle_{R_I}|\rangle_{N_F,T}$ but differs from the transition point identified by susceptibility $\chi$ at $\eta = 0.9\pi$. (h) Normalized TE versus $\langle|\langle A\rangle_{R_I}|\rangle_{N_F,T}$. (i)–(l) Transition between the aligned and chiral phases. (i) TE versus $w_{I\to F}$. The values for $w_{I\to F} < 0$ are significantly larger than those for $w_{I\to F} > 0$ because the rotation in the chiral phase creates large variations in orientations and thus enriches information content. (j) Total MI versus $w_{I\to F}$. (k) Normalized TE versus $w_{I\to F}$. The transition point is clearly at $w_{I\to F} = 0$, which agrees with the transition points identified by both $\langle|\langle A\rangle_{R_I}|\rangle_{N_F,T}$ and susceptibility $\chi$. (l) Normalized TE versus $\langle|\langle A\rangle_{R_I}|\rangle_{N_F,T}$.

The information transfer from influencer-neighbors to one particle for the aligned-disordered transition is shown in Figs. 11(a)–(d). As the noise strength $\eta$ increases, the TE increases rapidly initially, reaches a plateau value in the ordered states, decreases rapidly during the order-disorder transition, and becomes near zero in the disordered states [Fig. 11(a)]. Total MI decreases as $\eta$ increases [Fig. 11(b)]. The normalized TE increases slowly as $\eta$ increases in the ordered states, and decreases rapidly during the order-disorder transition, and becomes near zero in the disordered states [Fig. 11(c)]. Most notably, the normalized TE versus $\eta$ curve exhibits an inverted V-shape around the transition region, and the cusp point of the curve corresponds to the order-disorder transition. It suggests that the most critical role of interactions among individuals in driving collective phase transitions lies in the relative importance between individuals: the phase transition is associated with a sharp decrease in the amount of information transferred from the influencer's present to the follower's future, relative to the total amount of information transferred to the follower's future. This inverted V-shape of the normalized TE versus $\eta$ curve is similar to the shape of the influence $\langle|\langle A\rangle_{R_I}|\rangle_{N_F,T}$ versus $\eta$ curves [Fig. 10(b)]. The similarity suggests a direct relationship between normalized TE and the influence. We therefore plot these two quantities against each other in Fig. 11(d). The resulting curve showed two branches, which meet at a common peak at $\eta = 1.05\pi$. The upper branch corresponds to the ordered phase, and the lower branch corresponds to the disordered phase. Within each branch, normalized TE increases monotonically with influence, indicating that a stronger net directional pull from influencer neighbors consistently enhances the relative informational contribution of the influencer's present over the follower's own present. Interestingly, for the same level of influence, the disordered branch lies below the aligned branch. This is because strong noise suppresses information transfer from influencers to followers, as already evident in Fig. 11(a), where TE drops sharply in the disordered regime.

The information transfer from influencer neighbors to one follower for the chiral-disordered transition is shown in Figs. 11(e)–(h). As $\eta$ increases, the TE first rises slightly and then declines monotonically, eventually approaching zero in the fully disordered state [Fig. 11(e)]. The total MI decreases as $\eta$ increases [Fig. 11(f)]. The normalized TE increases slowly as $\eta$ increases in the ordered states, and decreases rapidly during the order-disorder transition, and becomes near zero in the disordered states [Fig. 11(g)]. The transition point identified by the peak in normalized TE coincides with that indicated by the influence measure in Fig. 10(d). The direct correlation between these two quantities is further illustrated in Fig. 11(h), which shows two branches that meet at a common peak at $\eta = 0.5\pi$ . The right branch corresponds to the chiral phase, whereas the left branch corresponds to the disordered phase. In the left branch, the normalized TE increases monotonically with influence, which is similar to Fig. 11(d) showing the transition process from the disordered side. In the right branch, the influence stays relatively constant while normalized TE increases. The shared peak of the two branches again emphasizes that the phase transition is related to a sharp drop in the relative importance of the influencer's present on the follower's future.

The information transfer from influencer neighbors to one follower for the aligned-chiral transition is shown in Figs. 11(i)–(l). The system is in the chiral phase for $w_{I\to F} < 0$ and in the aligned phase for $w_{I\to F} > 0$. As shown in Figs. 11(i) and 11(k), both the TE and the normalized TE are nearly constant within each phase, with significantly larger values in the chiral phase than in the aligned phase. Total MI almost remains unchanged across the transition [Fig. 11(j)]. Notably, the normalized TE also clearly identifies the transition point at $w_{I\to F} = 0$, which aligns with the behavior of the influence measure [Fig. 10(f)]. The relationship between normalized TE and influence is shown in Fig. 11(l), which also exhibits two branches that meet at $w_{I\to F} = 0$. The upper branch represents the chiral phase, while the lower branch represents the aligned phase. Unlike the branches induced by noise in Figs. 11(d) and 11(h), the branches in Fig. 11(l) arise from the different interaction patterns between the chiral and aligned phases. The chiral phase enriches angular variation, resulting in greater information transfer.

To summarize the key points in Figs. 10 and 11, we find that both normalized TE and influence $\langle|\langle A\rangle_{R_I}|\rangle_{N_F,T}$ peak at the same points in the two noise-induced transitions, and these points differ from the transition points identified by the traditional order parameters shown in Fig. 4. This finding suggests that noise-induced phase transitions are associated with changes in the relative importance of influencers' presents or followers' presents on followers' future, an aspect of collective dynamics not captured by the traditional order parameters. It reveals the underlying close connection between normalized TE and the influence, and it may be related to the hypothesis that a collective's ability to process information is maximum near its transition point [1,57,58].

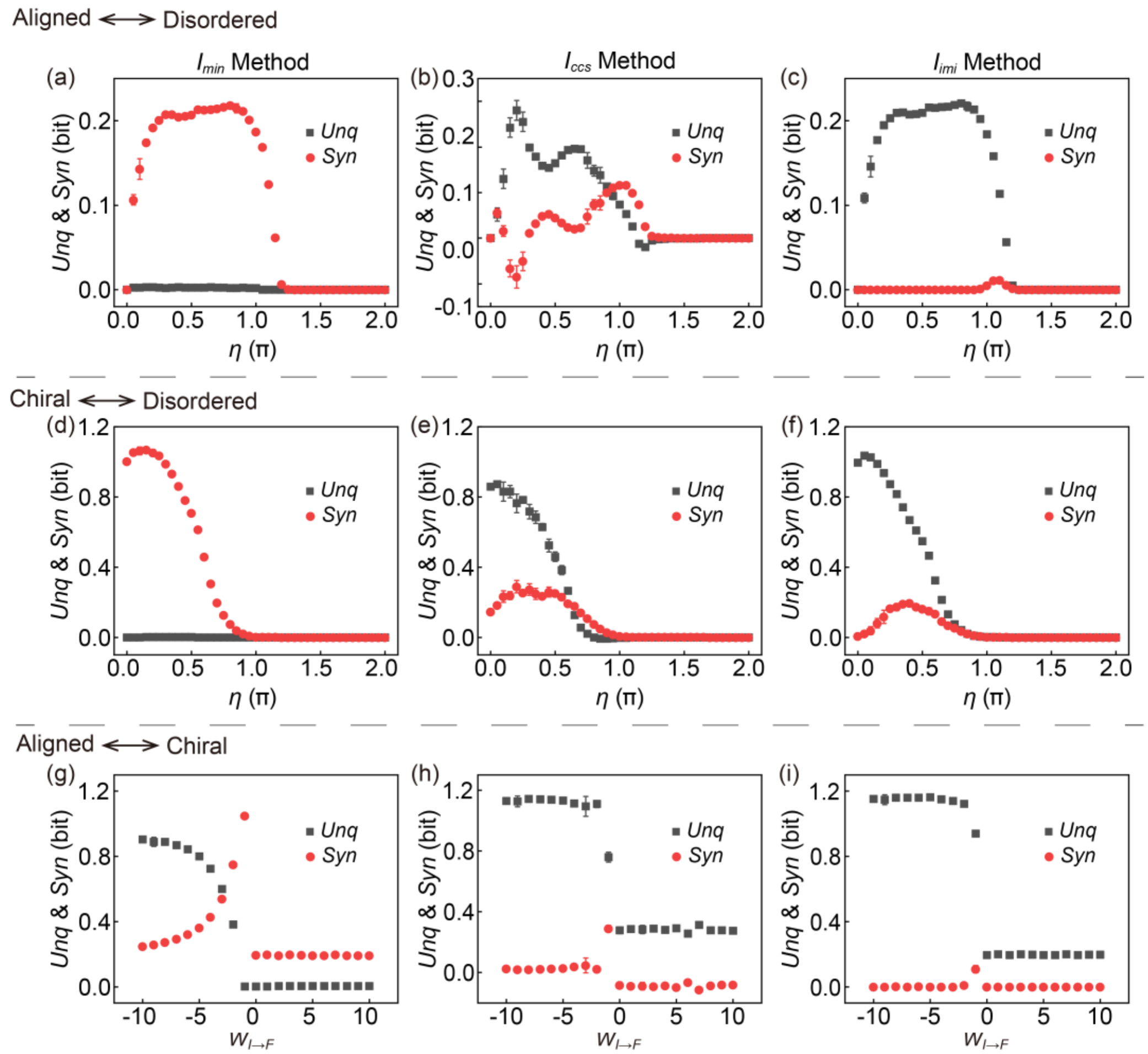

FIG. 12. Decomposition of TE for the phase transitions using three representative PID methods. The simulations are the same as the ones in Figs. 4, 10 and 11. (a)–(c) Transition between the aligned and disordered phases. (a) $I_{min}$ method. (b) $I_{ccs}$ method. (c) $I_{imi}$ method. $I_{min}$ attributes most of the TE to synergistic information, whereas $I_{ccs}$ and $I_{imi}$ attribute most of the TE to unique information. (d)–(f) Transition between the chiral and disordered phases. (d) $I_{min}$ method. (e) $I_{ccs}$ method. (f) $I_{imi}$ method. $I_{min}$ attributes most of the TE to synergistic information, whereas $I_{ccs}$ and $I_{imi}$ attribute most of the TE to unique information. (g)–(i) Transition between the aligned and chiral phases. (g) $I_{min}$ method. (h) $I_{ccs}$ method. (i) $I_{imi}$ method. $I_{min}$ attributes most of the TE to synergistic information for $w_{I \to F} > -2$ and to unique information for $w_{I \to F} < -2$, whereas $I_{ccs}$ and $I_{imi}$ attribute most of the TE to unique information. Although no universally accepted PID method currently exists, we find that the $I_{ccs}$ and $I_{imi}$ approaches are particularly suitable for analyzing information transfer in our system. They are computationally acceptable and, importantly, align with our physical intuition that influence should contain a significant unique information component, especially in the low-noise regime ($\eta$ small).

Finally, as an application, we employ our model to test several methods of PID. PID methods have been proposed to address the shortcomings of TE by decomposing it into unique (intrinsic) and synergistic information [59]. To provide some intuition on the PID methods, we evaluated seven PID methods using three binary test cases: XOR, OR, and ADD. For the XOR case, information is purely synergistic. For the OR case, all four components—unique, shared (redundant), and synergistic information—are expected to be non-zero. In the ADD (binary addition) case, only unique and synergistic components should be present, with no redundant contribution (see Appendix E for details). Three PID methods are selected here for further investigation in our collective dynamics study. The first is the original PID method proposed by Williams and Beer, which defines shared information via the average minimum specific information to complete the PID, abbreviated as the $I_{min}$ method [13]. The second method is the $I_{ccs}$ method, which measures redundant information based on pointwise changes in surprisal [36]; it can correctly distinguish each decomposed component in the three binary test cases. The third is the recently proposed $I_{imi}$ method, which measures the unique information by using the intrinsic mutual information to approximate the secret key agreement rate [16]. Due to the high computational cost of the $I_{ccs}$ method, calculations are performed for 10 randomly selected particles per simulation (compared to 100 for other methods); this sample size is sufficient for comparing $Unq$ and $Syn$ values as shown in Appendix E. Fig. 12 highlights the most interesting finding: the $I_{min}$ method attributes most of the TE to synergistic information ($Syn$), while the $I_{imi}$ and $I_{ccs}$ methods attribute the majority to unique information ($Unq$). This presents a fundamental question: which decomposition correctly reflects the information dynamics in our system?

We argue that the $I_{imi}$ and $I_{ccs}$ methods provide more appropriate decompositions for our system, based on the following reasoning. Our analysis in Fig. 11 reveals a link between the normalized TE and the influence $\langle|\langle A\rangle_{R_I}|\rangle_{N_F,T}$; we therefore investigate the nature of influence to uncover the nature of information transfer. According to Eq. (16), the influence term is the only term that contains neighbors' presents in calculating one particle's future state. Consequently, the nature of influence must inherently contain a unique information component. This element of uniqueness should dominate the regions of small noise. Both $I_{imi}$ and $I_{ccs}$ emphasize the unique information, especially in the low-noise, ordered regions, whereas $I_{min}$ contributes most of the information to the synergy. This conclusion is consistent with the previous findings that the $I_{min}$ method tends to overestimate synergistic information [14,15]. We note, however, that neither method is ideal. Based on binary cases and Fig. 12, $I_{imi}$ tends to underestimate synergistic information compared to $I_{ccs}$. Conversely, $I_{ccs}$ can yield negative values and is computationally more demanding than $I_{imi}$. Despite these limitations, both $I_{imi}$ and $I_{ccs}$ provide valuable and principled frameworks for analyzing information flow during phase transitions, particularly given the absence of a universally accepted PID method [4].

## VI. CONCLUSIONS

In conclusion, we have introduced an influence-based Vicsek model with non-reciprocal interactions. It clarifies the concept of influence, enables its quantification, and allows its direct comparison with information transfer calculated by various information-theoretic tools. Although no single universal relation between influence and information transfer exists, we have found several specific relations linking these two concepts at both the pairwise and collective levels.

At the pairwise level, we have found that at a fixed noise strength, the time average of the absolute values of pairwise influences, $\langle|A_{I\to F}|\rangle_T$, has a quasi-linear relation with the pairwise TE. At a fixed interaction weight, $\langle|A_{I\to F}|\rangle_T$ display a nonlinear relation with the pairwise TE because of the dual effects of noise: the noise of the influencer increases the information transfer to the follower, whereas the noise of the follower suppresses the information received by the follower.

At the collective level, we have found that $\langle|\langle A\rangle_{R_I}|\rangle_{N_F,T}$ (the absolute influencer-neighbor-averaged pairwise influences averaged over the number of followers and time) and normalized TE (calculated using an effective mean-field approximation) form two-branched relations that clearly identify the same transition points across three distinct phase transitions. These transition points differ from those indicated by classical order parameters. They reveal a different aspect of the collective dynamics: phase transitions are associated with changes in the relative importance of influencers' presents or followers' presents on followers' futures. Finally, we have tested three different PID methods and found that the interpretations given by the $I_{ccs}$ and $I_{imi}$ methods are most appropriate for the influence-based Vicsek model.

This work is the first step in differentiating the concept of influence from the concept of information transfer. Future work will extend it in several directions: conducting a comprehensive parameter scan to map out the phase diagram of our modified Vicsek model [47] and analyzing the relations between influence and information transfer on this phase diagram, exploring quantitative measures of influence in other systems, testing the generalizability of the various quantitative relations discovered in the influence-based Vicsek model, and examining the connections to standard thermodynamic quantities. Other techniques, like those employed in network science, will be explored to investigate collective behavior [60–68] and analyze how the influence and information organize collective dynamics. In addition, we will explore experimental systems based on programmable active matter [69], where we can measure both influence and information transfer directly.

## ACKNOWLEDGMENTS

This work was supported by the National Natural Science Foundation of China (project number 22175115), the Science and Technology Commission of Shanghai Municipality (project number 23ZR1433700), the start-up fund of GC-SJTU, and the Young Top-notch Talent Support Program.

## APPENDIX A: FURTHER DESCRIPTIONS ABOUT METHODS OF PARTIAL INFORMATION DECOMPOSITION

In Sec. II.B, we have introduced the basic framework of partial information decomposition (PID). A scheme of PID is shown in Fig. 2. Here, in Appendix A, we give brief descriptions of the decomposition methods used in the study. To make our description more general, we use sources $S_1$ and $S_2$ to represent $X(t)$ and $Y(t)$, respectively, and use target $T$ to represent $Y(t+\Delta t)$. We reproduce the Eqs. (8) to (10) below for the completeness of our description:

$$I(S_1, S_2; T) = Unq(S_1) + Unq(S_2) + Shd + Syn, \tag{A1}$$

$$I(S_1; T) = Unq(S_1) + Shd, \tag{A2}$$

$$I(S_2; T) = Unq(S_2) + Shd. \tag{A3}$$

Williams and Beer [13] proposed a measure of shared information based on the concept of specific information [70]. The specific information quantifies the information a source provides about a particular outcome. For example, the specific information about $T = t$ from source $S_1$ is defined as

$$I(T = t; S_1) = \sum_{s_1} p(s_1|t) \log \frac{p(t, s_1)}{p(t)p(s_1)}. \tag{A4}$$

The shared information is the expected value of the minimum specific information from both sources:

$$Shd = I_{min}(T; S_1, S_2) = \sum_{t} p(t) \min\{I(T = t; S_1), I(T = t; S_2)\}. \tag{A5}$$

This method is denoted as $I_{min}$ in the literature. We note that the recent work on Synergistic-Unique-Redundant Decomposition of causality (SURD) also uses the concept of specific information as its starting point [17]. For the case of two source variables, the SURD method is the same as $I_{min}$.

Harder et al. [14] proposed a measure of shared (redundant) information based on the concept of information projection in information geometry. The central idea is to project the conditional probability of one source variable, $p(t|s_1)$ or $p(t|s_1)$, onto an optimized probability distribution constructed from another source variable. Hence this method is called $I_{proj}$ in the literature. The resulting projected information $I_Z^\pi(X \searrow Y)$ and $I_Z^\pi(Y \searrow X)$ are used to construct redundant information as follows:

$$Shd = \min\{I_Z^\pi(X \searrow Y), I_Z^\pi(Y \searrow Z)\}. \tag{A6}$$

Bertschinger et al. [71] proposed a measure of unique information based on their intuition that unique and shared information should only depend on the marginal distributions of the target and individual sources, i.e., $p(t, s_1)$ and $p(t, s_2)$. This method is called $I_{broja}$ in the literature. They define unique information as follows:

$$Unq(S_1) = \min_{Q \in \Delta_p} I_Q(T; S_1|S_2), \tag{A7}$$

where $\Delta_p$ is the set of all joint distributions that have the same marginal distributions of the target and individual sources as the original joint distribution. $Q = q(t, s_1, s_2)$ is an element of $\Delta_p$, so $q(t, s_1) = p(t, s_1)$ and $q(t, s_2) = p(t, s_2)$. The subscript $Q$ in $I_Q$ means that the conditional mutual information $I_Q$ is calculated with respect to the joint distribution $Q$.

James et al. [72] and Kay et al. [73] independently proposed a measure of unique information based on the dependency lattice of maximum entropy models. This method is called $I_{dep}$ in the literature. The basic idea is that individual distributions of target, $p(t)$, and sources, $p(s_1)$ and $p(s_2)$, have the least amount of information, and the joint distributions $p(t, s_1, s_2)$ has the most amount of information. In between the two extremes, we have joint distributions that have an intermediate amount of information. These distributions form a dependency lattice of models. Unique information $Unq(S_1)$ is the minimum increase in information on lattice edges where $S_1$ and $T$ are combined to form a joint distribution.

Ince [36] proposed a measure of shared information based on pointwise change in surprisal. The term pointwise means that each realization of random variables is considered separately. They call their measure common change in surprisal and denote it using $I_{ccs}$. Surprisal of a realization $T = t$ is $-\log p(t)$, and the change in surprisal of the realization $T = t$ given $S_1 = s_1$ is $\log p(t|s_1)/p(t)$. The common change in surprisal considers the changes of surprisal given $S_1 = s_1$, $S_2 = s_2$, and $\{S_1 = s_1, S_2 = s_2\}$, as well as coinformation $c(t: s_1: s_2)$. If these four quantities are of the same sign, they contribute to shared information.

Finn and Lizier [74] proposed a measure of shared information based on pointwise partial information decomposition. This method is abbreviated as $I_{pm}$ in the literature. For each realization, the pointwise shared information is the difference between an informative part and a misinformative part. The informative part is $Shd^+ = \min\{h(s_1), h(s_2)\}$, where $h(s_i) = -\log p(s_i)$, with $i = 1, 2$ .The misinformative part is $Shd^- = \min\{h(s_1|t), h(s_2|t)\}$, with $h(s_i|t) = -\log p(s_i|t)$. Hence the pointwise shared information is $Shd^p = Shd^+ - Shd^-$. The shared information is then the average of pointwise shared information $Shd = \langle Shd^p \rangle$.

James et al. proposed a measure of unique information based on the idea of secret key agreement rate in cryptography [16,59]. This method is abbreviated as $I_{imi}$ in literature. Secret key agreement rate quantifies the rate of secret (information) shared between one source and the target, while the other source cannot know the secret. The secret agreement rate is not computable, but its upper bound, intrinsic mutual information (IMI), is computable. The unique information from the source $S_1$ to the target $T$ can be expressed as IMI as follows:

$$Unq(S_1) = I(T; S_1 \downarrow S_2) = \min_{p(\bar{s}_2|s_2)} I(T; S_1|\bar{S}_2), \quad \text{(A8)}$$

where $\bar{S}_2$ is an arbitrary stochastic function of $S_2$, and $p(\bar{s}_2|s_2)$ is the conditional probability. Figuratively speaking, $S_2$ is the eavesdropper who tries to steal the secret from $T$ and $S_1$, and IMI is the secret (unique information) between $T$ and $S_1$, which $S_2$ does not know. $\bar{S}_2$ represents any local modification of $S_2$ and is characterized by the conditional property $p(\bar{s}_2|s_2)$.

## APPENDIX B: PARAMETERS INVESTIGATION OF THE MODIFIED VICSEK MODEL

In this section, we explore some parameters of the modified Vicsek model to verify its effectiveness. By setting all the interaction weights to one, $R = 1$, and $\rho = 4$, Fig. 13 illustrates the evolution of the collective system via the mean speed $v_a(t)$ for different $\eta$. The data demonstrate that systems reach a stationary state after approximately 500 time steps, beyond which $v_a(t)$ fluctuates within a certain range. A larger interaction radius $R$ accelerates this convergence to stationarity. Based on the observation in Fig. 13, we therefore sample all data after the first 500 time steps for each simulation.

We scan the parameter $N$ in the simulations presented in Fig. 14. Fig. 14(a) shows that the aligned-disorder phase transition is preserved, consistent with the original Vicsek model [33]. The Binder cumulant [Fig. 14(b)] and the susceptibility [Fig. 14(c)] likewise exhibit the expected dependence on angular noise [45]. The $\langle |\langle A \rangle_{R_I}| \rangle_{N_F,T}$ are further tested here in Fig. 14(d). Its characteristic inverted-V shape persists and clearly identifies the transition point. Interestingly, for a fixed $\rho$, the curves for different $N$ collapse onto one curve, especially when $\eta < \pi$.We attribute this collapse to the ordered-state assumption that particles are mutually aligned; under this condition, the system size $N$ affects only the precision of statistical averages (over particles and time) but not the underlying mean behavior. Consequently, even a modest system size (e.g., $N = 40$) provides sufficient averaging, and all curves converge to the same universal form. Guided by the observations in Fig. 13 and taking computational cost into account, we select $N = 400$ to investigate the collective behavior in the main text.

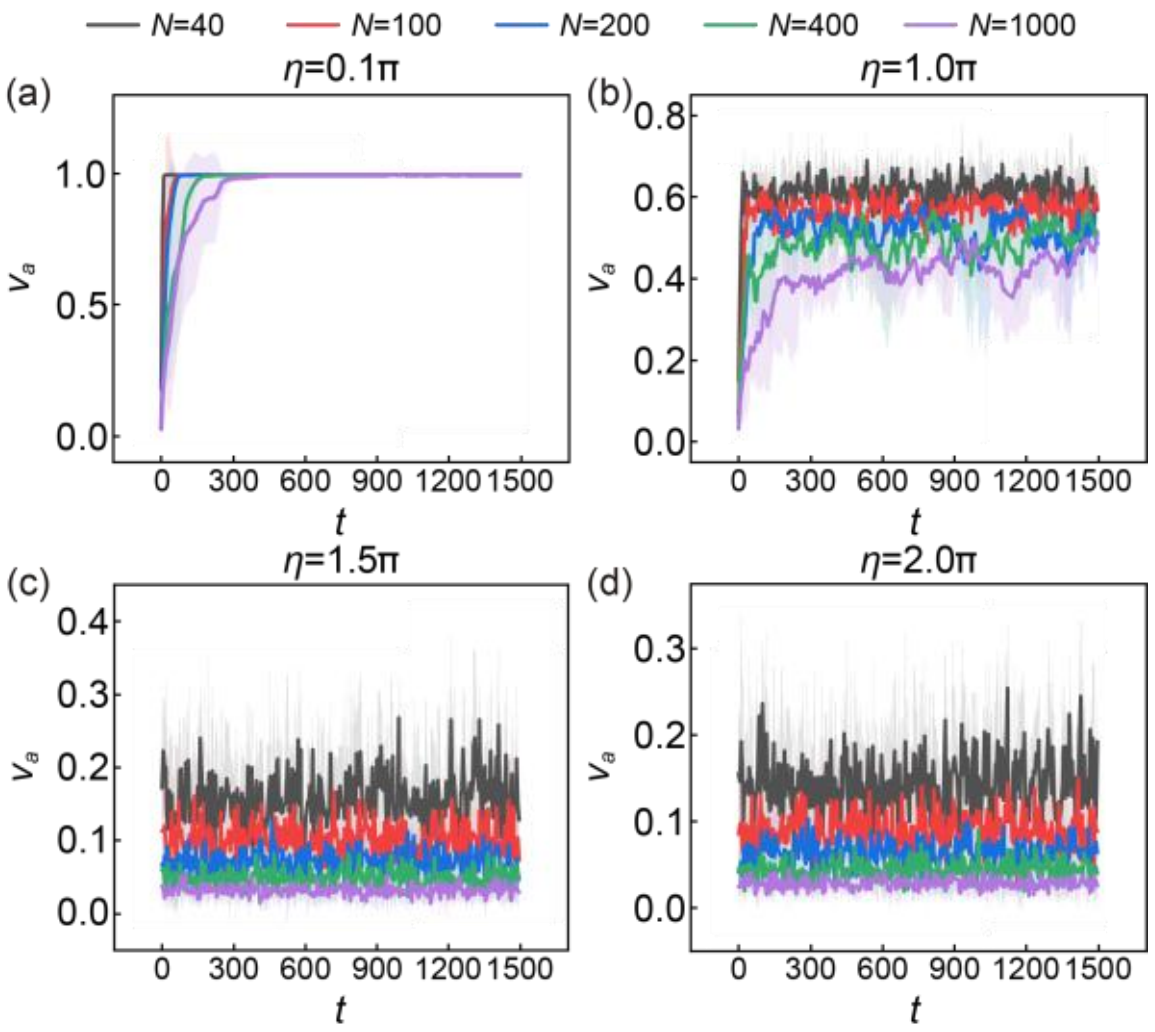

FIG. 13. $v_a(t)$ versus $t$ for different $N$ and $\eta$ with $R = 1$ and $\rho = 4$. The shaded area represents the standard deviation across five independent simulation runs. (a) $\eta = 0.1\pi$. (b) $\eta = 1.0\pi$. (c) $\eta = 1.5\pi$. (d) $\eta = 2.0\pi$. $v_a(t)$ reaches steady states after 500 time steps.

Examining the locations of these peaks in Figs. 14(c) and 14(d) reveals a difference [Fig. 15(a)]: the peaks in $\langle |\langle A \rangle_{R_I}| \rangle_{N_F,T}$ remain constant at $\eta = \pi$, independent of system size. In contrast, the peaks in the susceptibility $\mathcal{X}$ shift with $N$, moving towards $\eta = \pi$ as $N$ increases. This finite-size scaling of the peak positions of susceptibility is similar to standard critical behavior [75]. In standard critical phenomena, the peak of the susceptibility scales with the linear system size $L$ as $\chi_{max} \sim L^{\gamma/\nu}$, where $\chi_{max}$ is the peak value of $\chi$, $\gamma$ is the susceptibility critical exponent, and $\nu$ is the correlation length critical exponent [75]. For our system, the particle density is fixed at $\rho = 4$, so the linear size is $L = \sqrt{N/\rho}$. Fig. 14(c) provides the susceptibility curves for $N = 40, 100, 200, 400, 1000$ (corresponding to $L \approx 3.16, 5.00, 7.07, 10.00, 15.81$). We extract the peak value $\chi_{max}$ for each system size and perform a linear fitting on the log-log plot of $\chi_{max}$ versus $L$ as shown in Fig. 15(b). The fitted slope yields $\gamma/\nu = 1.70 \pm 0.04$. Moreover, the peak value for $\langle |\langle A \rangle_{R_I}| \rangle_{N_F,T}$ remains approximately 0.37 for all system sizes, which reinforces the interpretation that the influence-based order parameter $\langle |\langle A \rangle_{R_I}| \rangle_{N_F,T}$ exhibits a qualitatively different behavior (size-independent peak position) from the classical susceptibility.

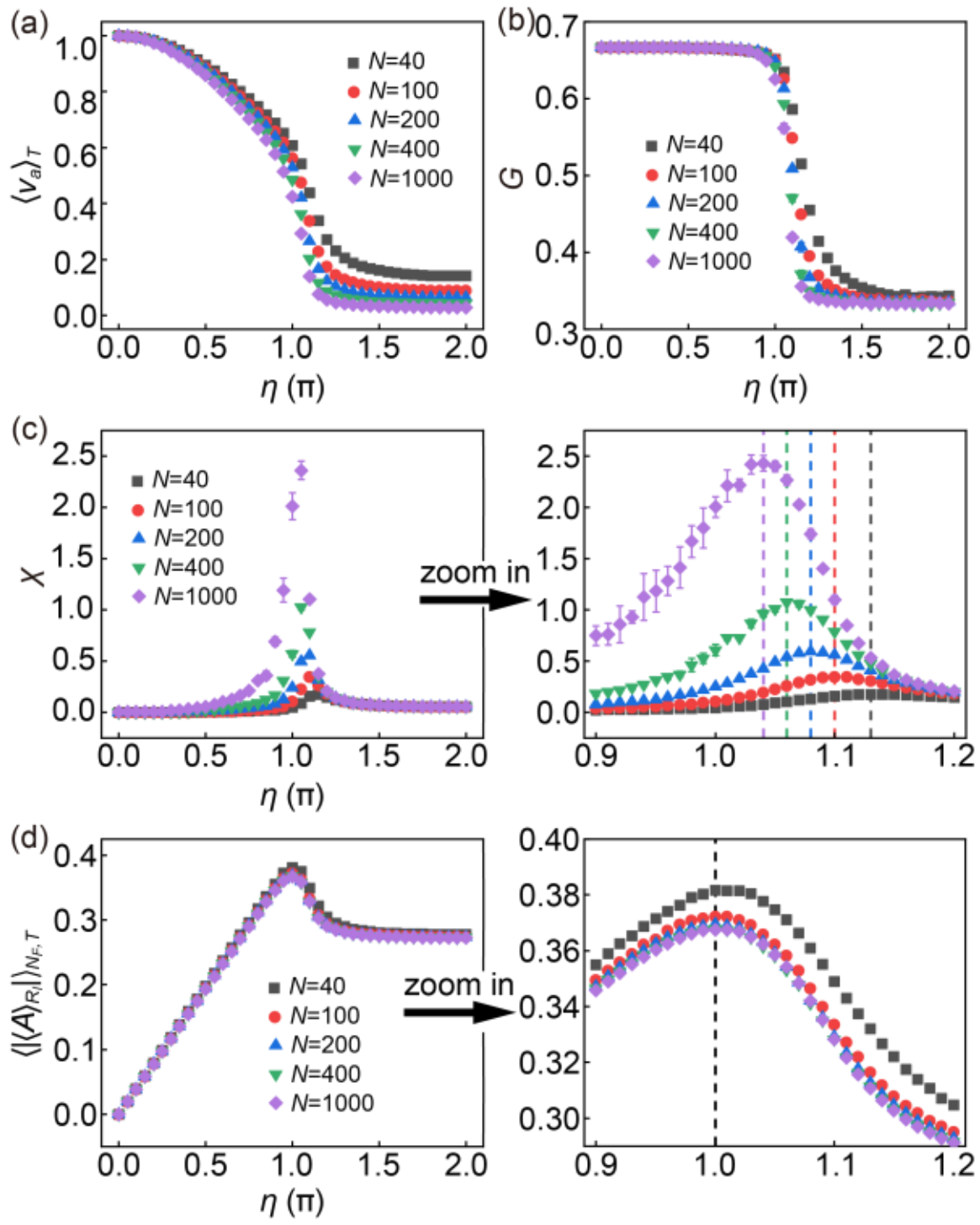

FIG. 14. Impact of model parameters $N$ on the characterizations of phase transitions using classical and influence-based order parameters with $R = 1$ and $\rho = 4$. (a) $\langle v_a \rangle_T$ versus $\eta$. (b) Binder cumulant $G$ versus $\eta$. The absence of a sharp drop in $G$ to negative values indicates that this aligned-to-disordered transition shows the behavior of a continuous-like transition at these finite system sizes. (c) Susceptibility $\mathcal{X}$ versus $\eta$. The peak position decreases with increasing $N$. (d) $\langle|\langle A\rangle_{R_I}|\rangle_{N_F,T}$ versus $\eta$. Different $N$ shares the same peak position at $\eta = \pi$.

Moreover, we investigate the parameter $R$ using $\langle v_a \rangle_T$ and $\langle|\langle A\rangle_{R_I}|\rangle_{N_F,T}$ in Fig. 16. Fig. 16(a) shows that increasing the radius of interaction $R$ can preserve and even sharpen the order-disorder transition. Fig. 16(b) shows that the peak location of the susceptibility $\chi$ does not decrease with increasing $R$; instead, it shifts to slightly higher noise values. This shift indicates that a larger number of interacting neighbors enhances the system's resilience to disorder. The absence of a sharp drop in $G$ to negative values for all $R$ indicates that the aligned-to-disordered transition appears continuous for the system sizes considered, and increasing $R$ does not alter the order of the transition, at least within the range of system sizes explored [Fig. 16(c)]. In Fig. 16(d), we further validate our analytic results for $\langle|\langle A\rangle_{R_I}|\rangle_{N_F,T}$. For $\eta < \pi$, data of different $R$ collapse onto one curve, consistent with the analytical prediction shown in Fig. 10(b). The asymptotic values as $\eta \to 2\pi$ are 0.273, 0.143, 0.096, 0.072, and 0.058 for $R = 1, 2, 3, 4, 5$, respectively. These values agree

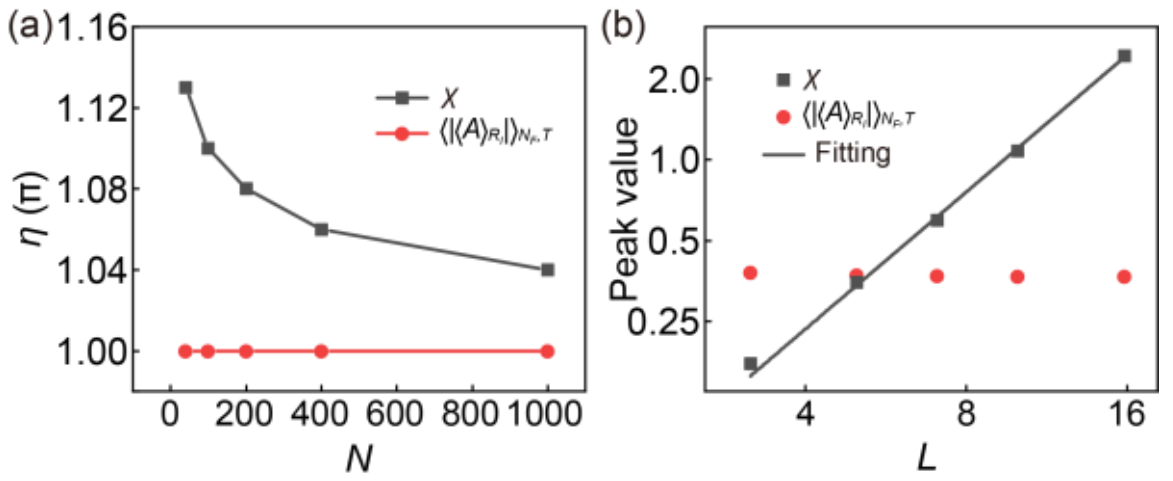

FIG. 15. Effects of the system size $N$ on the transition points. (a) The peak positions at $\eta$ versus $N$ for $\mathcal{X}$ and $\langle|\langle A\rangle_{R_I}|\rangle_{N_F,T}$. The peak position for $\chi$ decreases with increasing $N$, whereas the peak position for $\langle|\langle A\rangle_{R_I}|\rangle_{N_F,T}$ remains constant at $\eta = \pi$ across all system sizes. (b) Log-log plot of the peak value as a function of system size $L$ for $\mathcal{X}$ and $\langle|\langle A\rangle_{R_I}|\rangle_{N_F,T}$. The peak value for $\chi$ increases with increasing $L$, whereas the peak value for $\langle|\langle A\rangle_{R_I}|\rangle_{N_F,T}$ remains approximately 0.37 for all system sizes. A linear fit yields a slope of $1.70 \pm 0.04$, giving a ratio of the critical exponents $\gamma/\nu = 1.70$.

well with the analytical values that, for $R = 1, 2, 3, 4,$ and $5$, $\langle|\langle A\rangle_{R_I}|\rangle_{N_F,T} = 0.287, 0.144, 0.096, 0.072$, and $0.058$, respectively. We note, however, that a very small $R$ can suppress the emergence of a system-wide chiral phase: when $R$ is small, followers tend to aggregate with the particles of their same type. This localized clustering reduces their sensitivity to the repulsive effect exerted by influencers that enter their interaction radius. For simplicity and clarity, we therefore select $R = 5$ as the standard value for analyzing the three transition types in the main text.

Furthermore, we examine the effect of density $\rho$ for a fixed system size ($L = 10$). Fig. 17(a) shows that at low noise (approximately $\eta < \pi$), increasing density increases $\langle v_a \rangle_T$ reflecting the growing robustness of the collective. However, when noise is large (approximately $\eta > \pi$), the interaction is totally dominated by the noise, and increasing the density does not reduce the disorder. When $\eta \approx \pi$, increasing density first decreases $\langle v_a \rangle_T$ and then increases it. When the density is low, particles rarely interact; the noise is sufficiently strong to overwhelm any persistent orientation, causing the particles to move randomly. In this regime, adding more particles merely increases the number of uncorrelated random walkers, contributing negligibly to collective order. As the density increases further, interactions become frequent enough that the noise—though still substantial—can no longer fully suppress the ordering tendency arising from particle alignment. Consequently, further increases in density begin to enhance the global order, reversing the initial downward trend. Thus, the non-monotonic dependence reflects a density-driven crossover from a noise-dominated to an interaction-dominated regime. Fig. 17(b) shows the susceptibility $\mathcal{X}$ as a function of $\eta$. The local peak around $\rho = 2$ relates to the crossover in Fig. 17(a). The behavior of Binder cumulant $G$ versus $\rho$ is shown as Fig. 17(c). Fig. 17(d) shows the Influence measure $\langle|\langle A\rangle_{R_I}|\rangle_{N_F,T}$ versus $\rho$. For all $\eta$ shown, this quantity initially increases with $\rho$ because more particles interact with each other, thereby increasing the influence that a follower receives from its influencer neighbors. In the ordered states ($\eta = 0.5\pi$ and $1.0\pi$), further increases in $\rho$ add more aligned influencers, which continue to raise the influence on the followers. In the disordered state ($\eta = 1.5\pi$), however, additional influencer neighbors are themselves randomly oriented; their contributions tend to cancel one another out, thereby reducing the net influence. Guided by the observations in Fig. 17 and following the choice made in the original Vicsek model [33], we select $\rho = 4$ to investigate the collective behavior in the main text.

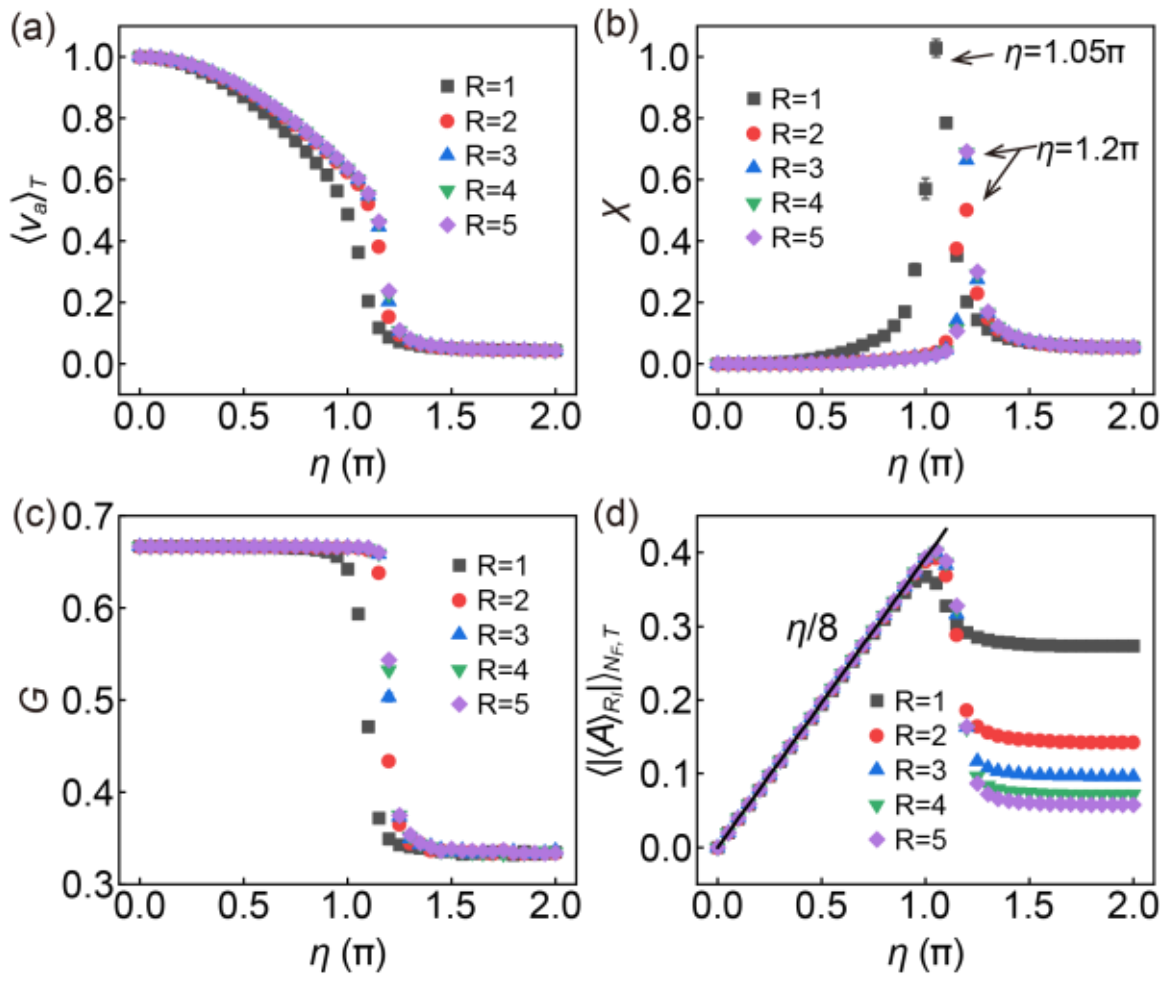

FIG. 16. Effects of neighborhood cutoff radius $R$ with $N = 400$ and $\rho = 4$. (a) $\langle v_a \rangle_T$ versus $\eta$. Increasing $R$ increasing the order of the aligned phases of the collective leading to larger $\langle v_a \rangle_T$. (b) Susceptibility $\mathcal{X}$ versus $\eta$. For $R = 1$, the peak is located at $\eta = 1.05\pi$; for all larger radii ($R = 2,3,4,5$), the peak shifts to $\eta = 1.2\pi$. The fact that the peak positions do not decrease with $R$ indicates that a larger number of interacting neighbors enhances the system's resilience to disorder. (c) Binder cumulant $G$ versus $\eta$. Increasing $R$ does not alter the order of the transition, at least within the range of system sizes explored. (d) $\langle|\langle A\rangle_{R_I}|\rangle_{N_F,T}$ versus $\eta$ for different $R$ with $N = 400$. For $\eta < \pi$, data of different $R$ collapse onto one line, $\langle|\langle A\rangle_{R_I}|\rangle_{N_F,T} = \eta/8$ [Fig. 10(b)], and the asymptotic values for $\eta \to 2\pi$ agree with the analytical result, $0.5\sqrt{2/(3R^2\rho_I)}$.

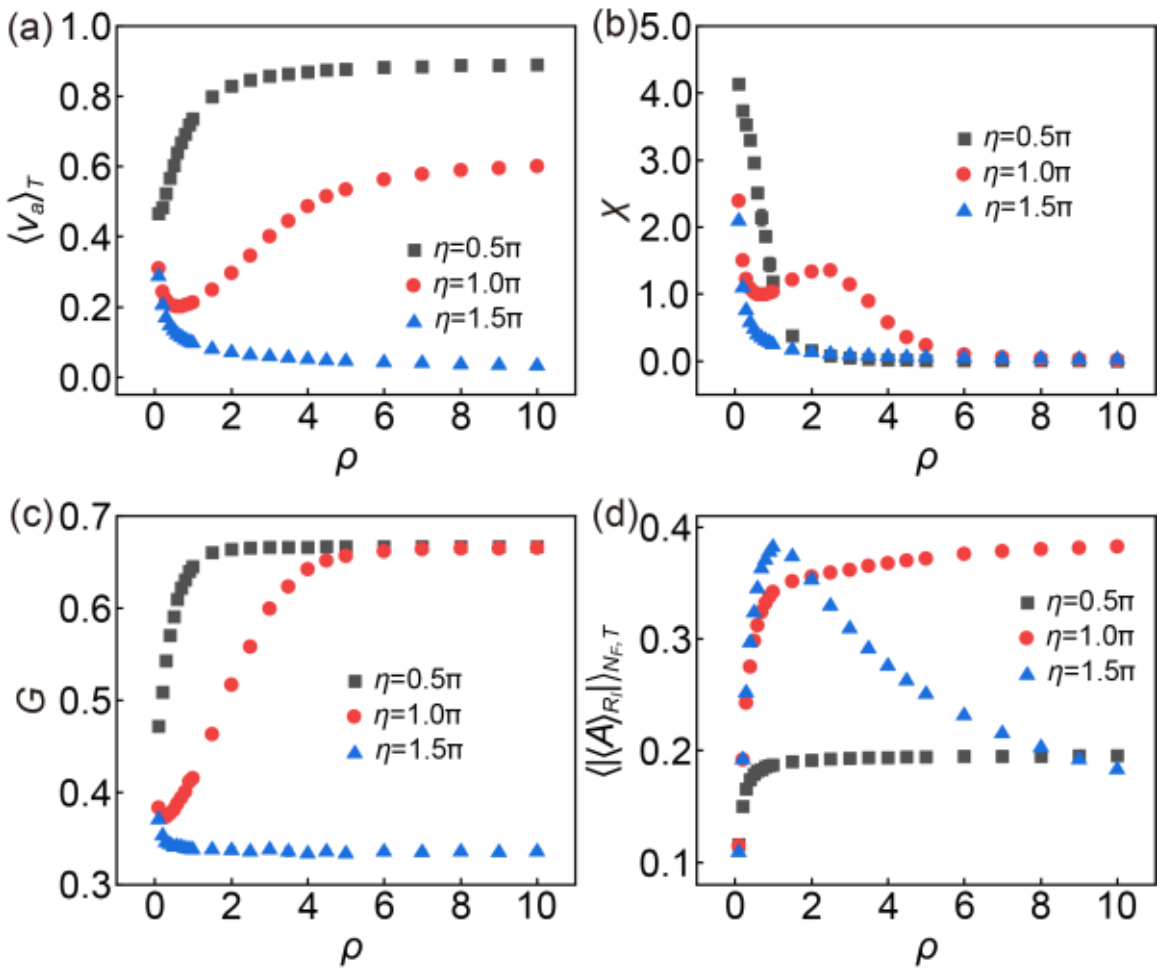

FIG. 17. Effects of densities $\rho$ with $R = 1$ and $L = 10$ for $\eta = 0.5\pi, 1.0\pi,$ and $1.5\pi$. (a) $\langle v_a \rangle_T$ versus $\rho$ for different $\eta$. For aligned phases ($\eta = 0.5\pi$), $\langle v_a \rangle_T$ increases with $\rho$; for $\eta = \pi$, increasing density first decreases $\langle v_a \rangle_T$ and then increases it. This non-monotonic reflects a density-driven crossover from a noise-dominated to an interaction-dominated regime; for disordered states ($\eta = 1.5\pi$), increasing density does not increase order. (b) Susceptibility $\mathcal{X}$ versus $\eta$. The local peak around $\rho = 2$ corresponds to the crossover observed in panel (a). (c) Binder cumulant $G$ versus $\rho$. (d) $\langle|\langle A\rangle_{R_I}|\rangle_{N_F,T}$ versus $\eta$. For all $\eta$ shown, this quantity initially increases with $\rho$ because more particles interact with each other, thereby increasing the influence that a follower receives from its influencer neighbors. The behavior upon further increasing $\rho$ differs depending on the system's state: ordered for $\eta = 0.5\pi$ and $1.0\pi$, and disordered for $\eta = 1.5\pi$ (see main text for details).

To briefly explore the existence of high-density traveling bands at a large system size with a small $R$, we set $R = 1$ and increase the arena size to $L = 200$ and the number of particles to $N = 20000$. Under these conditions, we indeed observe a traveling band for $\eta = 0.6\pi$, as shown in Fig. 18. This observed traveling band suggests that our modified Vicsek model may exhibit a first-order transition at a large system size [45]. We will explore this point in future work.

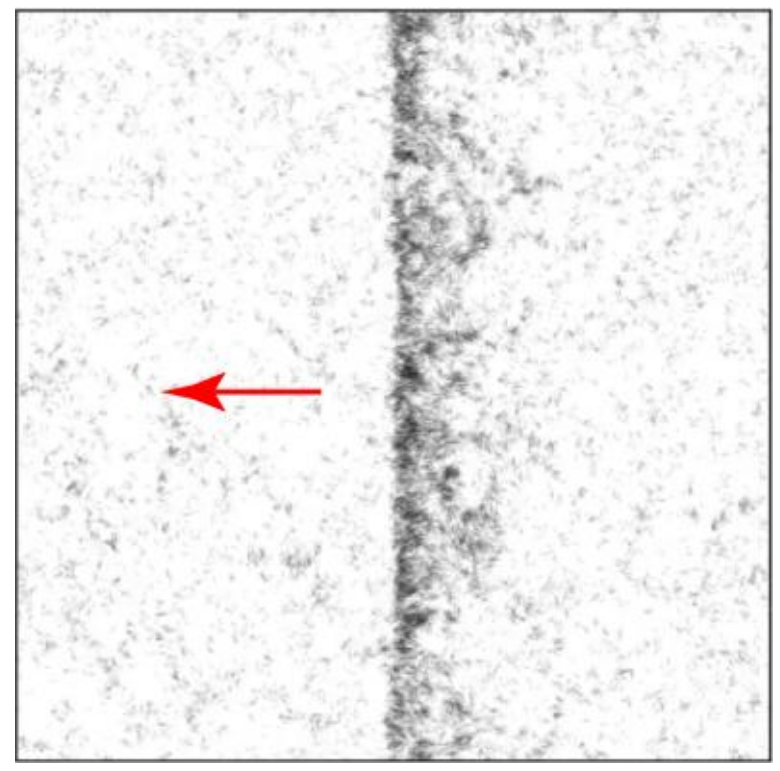

FIG. 18. Snapshot of the band for $N = 20000$, $L = 200$, $R = 1$, $v = 0.5$, and $\eta = 0.6\pi$.

## APPENDIX C: CALCULATION OF THE TIME-AVERAGED ABSOLUTE INFLUENCE IN PAIRWISE INTERACTIONS

The time-averaged absolute influence $\langle |A_{I\to F}| \rangle_T$ increases with $w_{I\to F}$ because the norm operation precedes the time averaging operation. Similarly, $\langle |A_{I\to F}| \rangle_T$ captures the effect of noise in the system and increases with noise strength $\eta$. To quantify the above qualitative statements, we derive an analytical expression for $\langle |A_{I\to F}| \rangle_T$.

**Case 1:**

When $w_{I\to F} > 0$, know that

$$A_{I\to F}(t) = \frac{w_{I\to F} F\big(\theta_I(t) - \theta_F(t)\big)}{w_{I\to F} + w_{F\to F}}, \quad \text{(C1)}$$

$$\theta_I(t) = F\big(\theta_I(t-\Delta t) + A_{F\to I}(t-\Delta t) + \beta_I(t-\Delta t)\big), \quad \text{(C2)}$$

and

$$\theta_F(t) = F\big(\theta_F(t-\Delta t) + A_{I\to F}(t-\Delta t) + \beta_F(t-\Delta t)\big). \quad \text{(C3)}$$

Assume that the influencer and the follower can agree on the same direction in the absence of noise at every time step:

$$\theta_I(t-\Delta t) + A_{F\to I}(t-\Delta t) = \theta_F(t-\Delta t) + A_{I\to F}(t-\Delta t). \quad \text{(C4)}$$

Therefore, the differences of $\theta_I(t)$ and $\theta_F(t)$ are the differences of $\beta_I(t-\Delta t)$ and $\beta_F(t-\Delta t)$:

$$A_{I\to F}(t) = \frac{w_{I\to F}}{w_{I\to F} + w_{F\to F}} \times F\big(\beta_I(t-\Delta t) - \beta_F(t-\Delta t)\big). \quad \text{(C5)}$$

$\langle |A_{I\to F}| \rangle_T$ is the average value of $|A_{I\to F}|$. As both $\beta_I(t-\Delta t)$ and $\beta_F(t-\Delta t)$ are uniformly distributed in $[-\eta/2, \eta/2]$, $\beta_I(t-\Delta t) - \beta_F(t-\Delta t)$ can be considered as the summation of two independent uniform distributions. Denote $F\big(\beta_I(t-\Delta t) - \beta_F(t-\Delta t)\big)$ as a random variable $Z$. Then calculating $\langle |A_{I\to F}| \rangle_T$ is to calculate the average value of $|Z|$. When $\eta \le \pi$,

$$Z = \beta_I(t-\Delta t) - \beta_F(t-\Delta t). \quad \text{(C6)}$$

The probability density function of $Z$ has two parts,

$$f_Z(z) = \begin{cases} \dfrac{1}{\eta^2}(z+\eta), & -\eta \le z < 0, \\ \dfrac{1}{\eta^2}(\eta - z), & 0 \le z \le \eta. \end{cases} \quad \text{(C7)}$$

The probability density function of $|Z|$ is

$$f_{|Z|}(|z|) = \frac{2}{\eta^2}(\eta - |z|), \quad 0 \le |z| \le \eta. \quad \text{(C8)}$$

Therefore, the average value of $|A_{I\to F}|$ when $\eta \le \pi$ is

$$\langle |A_{I\to F}| \rangle_T = \frac{w_{I\to F}}{w_{I\to F} + w_{F\to F}} \times \frac{\eta}{3}. \quad \text{(C9)}$$

When $\eta > \pi$, $F(\cdot)$ will fold the values of $\big(\beta_I(t-\Delta t) - \beta_F(t-\Delta t)\big)$ below $-\pi$ and above $+\pi$, so the probability density function of $Z$ has four parts,

$$f_Z(z) = \begin{cases} \dfrac{2}{\eta^2}(\eta - \pi), & -\pi \le z < -2\pi + \eta, \\ \dfrac{1}{\eta^2}(z+\eta), & -2\pi + \eta \le z < 0, \\ \dfrac{1}{\eta^2}(\eta - z), & 0 \le z < 2\pi - \eta, \\ \dfrac{2}{\eta^2}(\eta - \pi), & 2\pi - \eta \le z \le \pi. \end{cases} \quad \text{(C10)}$$

And the probability density function of $|Z|$ is

$$f_{|Z|}(|z|) = \begin{cases} \dfrac{2}{\eta^2}(\eta - |z|), & 0 \le |z| < 2\pi - \eta, \\ \dfrac{4}{\eta^2}(\eta - \pi), & 2\pi - \eta \le |z| \le \pi. \end{cases} \quad \text{(C11)}$$

Therefore,

$$\langle |A_{I\to F}| \rangle_T = \frac{w_{I\to F}}{w_{I\to F} + w_{F\to F}} \times \left(\frac{(2\pi - \eta)^3}{3\eta^2} + \frac{2\pi^2(\eta - \pi)}{\eta^2}\right) = \frac{w_{I\to F}}{w_{I\to F} + w_{F\to F}} \times \left(\frac{2\pi^3 - 6\pi^2\eta + 6\pi\eta^2 - \eta^3}{3\eta^2}\right). \quad \text{(C12)}$$

In conclusion

$$\langle |A_{I\to F}| \rangle_T = \frac{w_{I\to F}}{w_{I\to F} + w_{F\to F}} \times \begin{cases} \dfrac{\eta}{3}, & 0 \le \eta < \pi, \\ \dfrac{2\pi^3 - 6\pi^2\eta + 6\pi\eta^2 - \eta^3}{3\eta^2}, & \pi \le \eta \le 2\pi. \end{cases} \quad \text{(C13)}$$

It should be mentioned that, though two particles tend to align with each other, the assumption

$$\theta_I(t-\Delta t) + A_{F\to I}(t-\Delta t) = \theta_F(t-\Delta t) + A_{I\to F}(t-\Delta t). \tag{C14}$$

cannot be guaranteed for each step when $w_{I\to I} = w_{F\to I} = w_{F\to F} = 1$ and $w_{I\to F} \to \infty$. If we ignore the noise,

$$A_{I\to F} = \frac{w_{I\to F}F(\theta_I - \theta_F)}{w_{F\to F} + w_{I\to F}} = \frac{w_{I\to F}F(\theta_I - \theta_F)}{1 + w_{I\to F}} \approx \frac{w_{I\to F}(\theta_I - \theta_F)}{1 + w_{I\to F}} \tag{C15}$$

and

$$A_{F\to I} = \frac{w_{F\to I}F(\theta_F - \theta_I)}{w_{F\to I} + w_{I\to I}} = \frac{F(\theta_F - \theta_I)}{2} \approx \frac{\theta_F - \theta_I}{2}. \tag{C16}$$

Thus,

$$\theta_I + A_{F\to I} = \frac{\theta_F + \theta_I}{2} \tag{C17}$$

and

$$\theta_F + A_{I\to F} = \frac{\theta_F + w_{I\to F}\theta_F + w_{I\to F}(\theta_I - \theta_F)}{1 + w_{I\to F}} = \frac{\theta_F + w_{I\to F}\theta_I}{1 + w_{I\to F}}. \tag{C18}$$

The assumption

$$\theta_I + A_{F\to I} = \theta_F + A_{I\to F} \tag{C19}$$

holds for each step when $w_{I\to F} \to 1$. Otherwise, there will be an extra term in $A_{I\to F}(t)$ which means

$$A_{I\to F}(t) = \frac{w_{I\to F}}{w_{I\to F} + w_{F\to F}} \times F\left(\hat{\theta}(t-\Delta t) + \beta_I(t-\Delta t) - \beta_F(t-\Delta t)\right). \tag{C20}$$

$\hat{\theta}(t-\Delta t)$ represents the error after two particles undergo a single alignment.

$$\hat{\theta} = \frac{\theta_F + \theta_I}{2} - \frac{\theta_F + w_{I\to F}\theta_I}{1 + w_{I\to F}} = \frac{(\theta_I - \theta_F)(1 - w_{I\to F})}{2(1 + w_{I\to F})}. \tag{C21}$$

$\hat{\theta}$ goes to 0 if there is no noise. $\hat{\theta}$ will continuously affect $A_{I\to F}(t)$ when the environment is noisy. One can reduce the effect of $\hat{\theta}$ by setting $w_{F\to I} \to 0$ or $w_{I\to I} \to \infty$.

**Case 2:**

When $w_{I\to F} < 0$, the calculation becomes complex but still analytically tractable.

$$A_{I\to F}(t) = \frac{|w_{I\to F}|F(\theta_I(t) - \theta_F(t) + \pi)}{|w_{I\to F}| + |w_{F\to F}|}, \tag{C22}$$

$$\theta_I(t) = F\left(\theta_I(t-\Delta t) + A_{F\to F}(t-\Delta t) + \beta_I(t-\Delta t)\right), \tag{C23}$$

and

$$\theta_F(t) = F\left(\theta_F(t-\Delta t) + A_{I\to F}(t-\Delta t) + \beta_F(t-\Delta t)\right). \tag{C24}$$

Assume the follower is on the anti-clockwise side (same for clockwise) and denote $w_{I\to F}$ as $w$. Then,

$$A_{I\to F} = \frac{|w|F(\theta_I - \theta_F + \pi)}{1 + |w|} = \frac{|w|(\pi - \Delta\theta)}{1 + |w|} \tag{C25}$$

and

$$A_{F\to I} = \frac{F(\theta_F - \theta_I)}{2} = \frac{\Delta\theta}{2}. \tag{C26}$$

When the system reaches the stationary state

$$A_{I\to F} = A_{F\to I} \Rightarrow \Delta\theta = \frac{2|w|}{1 + 3|w|}\pi. \tag{C27}$$

Consider the noise

$$A_{I\to F}(t) = \frac{|w|}{1 + |w|}F\left(\pi - \Delta\theta + \beta_F(t-\Delta t) + \beta_I(t-\Delta t)\right) = \frac{|w|}{1 + |w|}F\left(\frac{1 + |w|}{1 + 3|w|}\pi + \beta_F(t-\Delta t) + \beta_I(t-\Delta t)\right). \tag{C28}$$

Denote $m = \frac{|w|}{1+|w|}$, $W = \frac{1+|w|}{1+3|w|}$, and $Z = F\left(W\pi + \beta_F(t-\Delta t) + \beta_I(t-\Delta t)\right)$. $Z$ is a random variable and

$$A_{I\to F}(t) = mZ. \tag{C29}$$

Then

$$\langle |A_{I\to F}(t)| \rangle_T = \langle m|Z| \rangle, \tag{C30}$$

where $f_Z(z)$ is probability density function of $S$. Firstly, discuss the $f_Z(z)$ in different $w$ and $\eta$.

**Case 2.1:** When $\eta = 0$,

$$f_Z(z) = \delta(z - W\pi), \tag{C31}$$

where $\delta$ denotes the Dirac delta function.

**Case 2.2:** When $0 < \eta \le \pi$,

Case 2.2.1: For $0 < \eta \le \pi - W\pi$,

$$f_Z(z) = \frac{\eta - |z - W\pi|}{\eta^2}, W\pi - \eta \le z \le W\pi + \eta. \tag{C32}$$

$$
\langle|A_{I\to F}(t)|\rangle_T =
\begin{cases}
mW\pi, & 0 \le \eta \le W\pi, \\
\frac{m}{3\eta^2}\left(-\pi^3 W^3 + 3\pi^2 W^2 \eta + \eta^3\right), & W\pi < \eta \le \pi - W\pi, \\
\frac{m\pi}{3\eta^2}\left(\pi^2\left(1 + W(-3 + (3 - 2W)W)\right) + 3\pi(-1 + 2W)\eta - 3(-1 + W)\eta^2\right), & \pi - W\pi < \eta \le W\pi + \pi, \\
\frac{m}{3\eta^2}\left(\pi^3(2 - (-6 + W)W^2) - 3\pi^2(2 + W^2)\eta + 6\pi\eta^2 - \eta^3\right), & W\pi + \pi \le \eta \le 2\pi - W\pi, \\
\frac{m\pi}{\eta^2}\left(\pi^2(-2 + 4W) + \pi(2 - 4W)\eta + W\eta^2\right), & 2\pi - W\pi < \eta \le 2\pi.
\end{cases}
\tag{C36}
$$

Case 2.2.2: For $\pi - W\pi < \eta \le \pi$,

$$
f_Z(z) =
\begin{cases}
\frac{\eta - z - 2\pi + W\pi}{\eta^2}, & -\pi \le z \le W\pi + \eta - 2\pi, \\
\frac{\eta - |z - W\pi|}{\eta^2}, & W\pi - \eta < z \le \pi, \\
0, & \text{otherwise.}
\end{cases}
\tag{C33}
$$

**Case 2.3:** When $\pi < \eta \le 2\pi$,

Case 2.3.1: For $\pi < \eta < W\pi + \pi$,

$$
f_Z(z) =
\begin{cases}
\frac{\eta - z - 2\pi + W\pi}{\eta^2}, & -\pi \le z \le W\pi - \eta, \\
\frac{2(\eta - \pi)}{\eta^2}, & W\pi - \eta < z \le W\pi + \eta - 2\pi, \\
\frac{\eta + z - W\pi}{\eta^2}, & W\pi + \eta - 2\pi < z \le W\pi, \\
\frac{\eta - z + W\pi}{\eta^2}, & W\pi < z \le \pi.
\end{cases}
\tag{C34}
$$

Case 2.3.2: For $W\pi + \pi \le \eta \le 2\pi$,

$$
f_Z(z) =
\begin{cases}
\frac{2(\eta - \pi)}{\eta^2}, & -\pi \le z \le W\pi + \eta - 2\pi, \\
\frac{\eta + z - W\pi}{\eta^2}, & W\pi + \eta - 2\pi < z \le W\pi, \\
\frac{\eta - z + W\pi}{\eta^2}, & W\pi < z \le W\pi + 2\pi - \eta, \\
\frac{2(\eta - \pi)}{\eta^2}, & W\pi + 2\pi - \eta < z \le \pi.
\end{cases}
\tag{C35}
$$

Fig. 19 illustrates the probability density function for $f_Z(z)$ in the four cases listed above. $W\pi + \beta_F(t - \Delta t) + \beta_I(t - \Delta t)$ is a triangular distribution whose average value is $W\pi$. Calculating $\langle|A_{I\to F}(t)|\rangle_T$ is folding and translating the triangular distribution and doing the average. Thus, the expression of $\langle|A_{I\to F}(t)|\rangle_T$ can be derived and is given by Eq. (24). For convenience, we restate it here as Eq. (C36). Some special cases are given as follows:

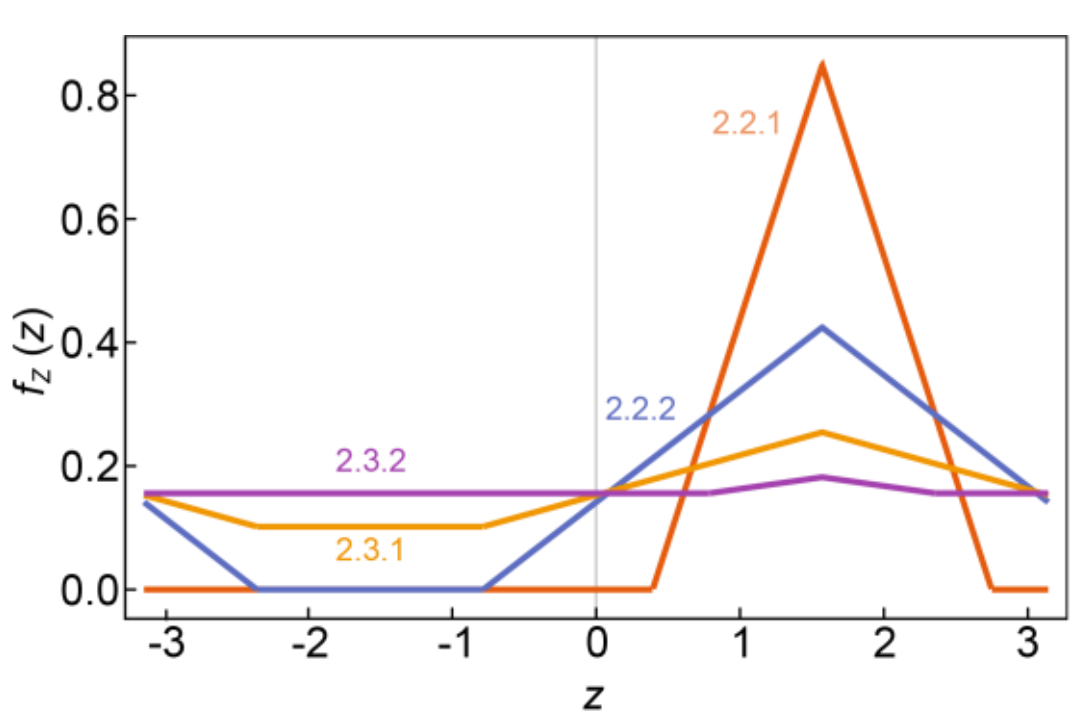

FIG. 19. The probability density function of $f_Z(z)$ for the cases 2.2.1, 2.2.2, 2.3.1, and 2.3.2, according to Eqs. (C32), (C33), (C34), and (C35).

**Special case 1:** When $w = -1$ and varying $\eta$, $W = \frac{1}{2}$ and $m = \frac{1}{2}$. One can obtain that

$$
\langle|A_{I\to F}(t)|\rangle_T = \langle m|Z|\rangle = \frac{\pi}{4}. \tag{C37}
$$

The time-averaged absolute influence is a constant with is consistent with the simulation results.

**Special case 2:** When $\eta = 0.2\pi = \frac{1}{5}\pi$ and varying $w$.

$$
f_Z(z) = \frac{\frac{1}{5}\pi - |s - W\pi|}{\frac{1}{25}\pi^2},
\quad s \in \left[W\pi - \frac{1}{5}\pi, W\pi + \frac{1}{5}\pi\right]. \tag{C38}
$$

As $W\pi - \frac{1}{5}\pi > 0$,

$$
\begin{aligned}
\langle|A_{I\to F}(t)|\rangle_T &= \langle m|Z|\rangle \\
&= \frac{|w|}{1 + |w|} \times \frac{1 + |w|}{1 + 3|w|}\pi \\
&= \frac{|w|}{1 + 3|w|}\pi.
\end{aligned}
\tag{C39}
$$

**Special case 3:** When $\eta = \pi$ and varying $w$. We can obtain

$$f_Z(z) = \begin{cases} \dfrac{-z + W\pi - \pi}{\eta^2}, & -\pi < z \le W\pi - \pi, \\ \dfrac{\pi + z - W\pi}{\eta^2} & W\pi - \pi < z \le W\pi, \\ \dfrac{\pi - z + W\pi}{\eta^2}, & W\pi < z \le \pi. \end{cases} \quad \text{(C40)}$$

Thus,

$$\langle |A_{I\to F}(t)| \rangle_T = \langle m|Z| \rangle = \frac{m\pi}{3}\left(1 + (3 - 2W)W^2\right). \quad \text{(C41)}$$

**Special case 4:** When $\eta = 2\pi$ and varying $w$.

$$f_Z(z) = \frac{2(\eta - \pi)}{\eta^2}, \quad -\pi < z \le \pi, \quad \text{(C42)}$$

and

$$\langle |A_{I\to F}(t)| \rangle_T = \frac{|w|}{1 + |w|} \times \frac{1}{2}\pi = \frac{|w|\pi}{2 + 2|w|}. \quad \text{(C43)}$$

## APPENDIX D: CALCULATION OF THE ABSOLUTE INFLUENCER-NEIGHBOR-AVERAGED PAIRWISE INFLUENCES AVERAGED OVER THE NUMBER OF FOLLOWERS AND TIME IN COLLECTIVE INTERACTIONS

**Case 1**: Transition between aligned and disordered phases

The absolute influencer-neighbor-averaged pairwise influences averaged over the number of followers and time, $\langle |\langle A \rangle_{R_I}| \rangle_{N_F,T}$, has a linear relationship with $\eta$ when the noise strength is small ($\eta < \pi$). This linear relationship is analyzed here.

First, we simplify the expression of $\langle A \rangle_{R_I}$ using $w_{j\to i} = 1$.

$$\begin{aligned} \langle A_i(t) \rangle_{R_I} &= \sum_{j\in\Omega_I} A_{j\to i}(t) \\ &= \sum_{j\in\Omega_I} \frac{w_{j\to i} F\left[\theta_j(t) - \theta_i(t)\right] s_{ij}(t)}{\sum_k w_{k\to i} s_{ik}(t)} \\ &= \sum_{j\in\Omega_I} \frac{F\left[\theta_j(t) - \theta_i(t)\right] s_{ij}(t)}{\sum_k s_{ik}(t)}. \end{aligned} \quad \text{(D1)}$$

Next, in the ordered state, all particles move in the same direction $\theta_{order}$. We assume that the neighbors' influences cause a particle to move in the direction $\theta_{order}$, which means

$$\theta_{order}(t - \Delta t) = \theta_i(t - \Delta t) + A_i(t - \Delta t). \quad \text{(D2)}$$

Therefore,

$$\begin{aligned} \theta_i(t) &= F\left(\theta_i(t - \Delta t) + A_i(t - \Delta t) + \beta_i(t - \Delta t)\right) \\ &= F\left(\theta_{order}(t - \Delta t) + \beta_i(t - \Delta t)\right). \end{aligned} \quad \text{(D3)}$$

Similarly,

$$\begin{aligned} \theta_j(t) &= F\left(\theta_j(t - \Delta t) + A_j(t - \Delta t) + \beta_j(t - \Delta t)\right) \\ &= F\left(\theta_{order}(t - \Delta t) + \beta_j(t - \Delta t)\right). \end{aligned} \quad \text{(D4)}$$

Therefore, the differences of $\theta_i(t)$ and $\theta_j(t)$ are caused by noise, which equals the differences of $\beta_i(t - \Delta t)$ and $\beta_j(t - \Delta t)$.

$$\begin{aligned} \langle A_i(t) \rangle_{R_I} &= \sum_{j\in\Omega_I} \frac{F\left[\theta_j(t) - \theta_i(t)\right] s_{ij}(t)}{\sum_k s_{ik}(t)} \\ &= \sum_{j\in\Omega_I} \frac{F\left[\beta_j(t - \Delta t) - \beta_i(t - \Delta t)\right] s_{ij}(t)}{\sum_k s_{ik}(t)}. \end{aligned} \quad \text{(D5)}$$

when $\eta < \pi$, so $\left[\beta_j(t - \Delta t) - \beta_i(t - \Delta t)\right] \in [-\pi, \pi]$, which means

$$F\left[\beta_j(t - \Delta t) - \beta_i(t - \Delta t)\right] = \beta_j(t - \Delta t) - \beta_i(t - \Delta t). \quad \text{(D6)}$$

Thus,

$$\begin{aligned} \langle A_i(t) \rangle_{R_I} &= \sum_{j\in\Omega_I} \frac{\left[\beta_j(t - \Delta t) - \beta_i(t - \Delta t)\right] s_{ij}(t)}{\sum_k s_{ik}(t)} \\ &= \sum_j \frac{\beta_j(t - \Delta t) s_{ij}(t)}{\sum_k s_{ik}(t)} - \frac{n_I}{n_I + n_F} \beta_i(t - \Delta t). \end{aligned} \quad \text{(D7)}$$

By setting an equal number of influencers and followers, we assume the number of neighbor influencers is equal to the number of neighbor followers, i.e., $n_I = n_F$. The first term can be assumed to be zero when there are enough neighbors, i.e., when the density is high enough, as the mean value of $\beta_j$ is zero. Therefore,

$$\langle A_i(t) \rangle_R = -\frac{1}{2} \beta_i(t - \Delta t). \quad \text{(D8)}$$

Because $\beta_i(t - \Delta t)$ is uniformly distributed in $\left[-\frac{\eta}{2}, \frac{\eta}{2}\right]$, its norm $|\beta_i(t - \Delta t)|$ is uniformly distributed in $[0, \eta/2]$. Therefore, the average of $|\beta_i(t - \Delta t)|$ is $\eta/4$:

$$\begin{aligned} \langle |\langle A \rangle_{R_I}| \rangle_{N_F,T} &= \frac{1}{T} \sum_{t=1}^{T} \frac{1}{N_F} \sum_{i\in\Omega_F} \left| \sum_{j\in\Omega_I} A_{j\to i}(t) \right| \\ &= \frac{1}{T} \sum_t \frac{1}{N} \sum_i \left| \frac{1}{2} \beta_i(t - \Delta t) \right| = \frac{\eta}{8}. \end{aligned} \quad \text{(D9)}$$

The case for disordered states is more difficult to analyze. Qualitatively, the decrease in $\langle |\langle A \rangle_{R_I}| \rangle_{N_F,T}$ can be explained as follows. Given angular parameters

$\theta_i(t) \in [0, 2\pi]$ and $\theta_j(t) \in [0, 2\pi]$, their difference $\left(\theta_j(t) - \theta_i(t)\right) \in (-2\pi, 2\pi)$. $F(\cdot)$ maps this difference to the $(-\pi, \pi)$ interval. This mapping reduces the value of effective influence, as $\left(\theta_j(t) - \theta_i(t)\right)$ often exceeds the $(-\pi, \pi)$ bounds. Therefore, $F(\cdot)$ depresses the effect of noise and converts large differences into small differences. This mapping in $F(\cdot)$ is the reason why$\langle|\langle A\rangle_{R_I}|\rangle_{N_F,T}$ begins to decrease.

The asymptotic value of $\langle|\langle A\rangle_{R_I}|\rangle_{N_F,T}$ at $\eta = 2\pi$ can be derived as follows. In

$$\sum_{j\in\Omega_I} A_{j\to i}(t) = \sum_{j\in\Omega_I} \frac{F\left[\theta_j(t) - \theta_i(t)\right] s_{ij}(t)}{\sum_k s_{ik}(t)}, \quad \text{(D10)}$$

because the noise is at the maximum value $\eta = 2\pi$, both $\theta_j(t)$ and $\theta_i(t)$ are independently and uniformly distributed over $[-\pi, \pi]$. Consequently, $F\left[\theta_j(t) - \theta_i(t)\right]$, after folding into the range $(-\pi, \pi]$, is also uniformly distributed $(-\pi, \pi]$. The variance of a uniform distribution over $(-\pi, \pi]$ is $\pi^2/3$, so, according to the central limit theory

$$\langle A\rangle_{R_I} = \sum_{j\in\Omega_I} A_{j\to i}(t) \sim \frac{1}{2}\mathcal{N}\left(0, \frac{\pi^2}{3n_I}\right). \quad \text{(D11)}$$

Here $\mathcal{N}\left(0, \frac{\pi^2}{3n_I}\right)$ denotes a Gaussian distribution with mean 0 and variance $\frac{\pi^2}{3n_I}$. Consequently, $\langle|\langle A\rangle_{R_I}|\rangle_{N_F,T}$ corresponds to the expected absolute value of a Gaussian random variable. Since averaging over $N$ particles and over time series provides sufficient statistical sampling, $\langle|\langle A\rangle_{R_I}|\rangle_{N_F,T}$ is the expected value of $\left|\langle A\rangle_{R_I}\right|$ over a Gaussian distribution $\frac{1}{2}\mathcal{N}\left(0, \frac{\pi^2}{3n_I}\right)$.

$$\begin{aligned}
\langle|\langle A\rangle_{R_I}|\rangle_{N_F,T} &= \left\langle\left|\frac{x}{2}\right|\right\rangle_{N_F,T} \approx E\left[\left|\frac{x}{2}\right|\right] \\
&= \int_{-\pi}^{\pi} \frac{\left|\frac{x}{2}\right|}{\sqrt{\frac{2\pi^3}{3n_I}}} \exp\left(-\frac{x^2}{\frac{2\pi^2}{3n_I}}\right) dx \\
&= \frac{1}{2}\sqrt{\frac{2\pi}{3n_I}} \approx \frac{1}{2}\sqrt{\frac{2\pi}{3\pi R^2 \rho_I}} \\
&= \frac{1}{2}\sqrt{\frac{2}{3R^2\rho_I}}. \quad \text{(D12)}
\end{aligned}$$

For $\rho = 4$, $\rho_I = 2$, $R = 5$ and $N = 400$, $\langle|\langle A\rangle_{R_I}|\rangle_{N_F,T} \approx 0.058$ which shows excellent agreement with the simulation result, $\langle|\langle A\rangle_{R_I}|\rangle_{N_F,T} \approx 0.058$.

**Case 2**: Transition between chiral and disordered phases

Denote $w_{I\to F}$ as $w$. We consider the averaged influence the influencers give to the followers $\langle|\langle A\rangle_{R_I}|\rangle_{N_F,T}$, where

$$\begin{aligned}
\langle A_i(t)\rangle_{R_I} &= \sum_{j\in\Omega_I} A_{j\to i}(t) \\
&= \sum_{j\in\Omega_I} \frac{|w_{j\to i}| F\left[\theta_j(t) - \theta_i(t) + \pi\right] s_{ij}(t)}{\sum_k |w_{k\to i}| s_{ik}(t)}, \\
& \qquad i \in \Omega_F. \quad \text{(D13)}
\end{aligned}$$

Assume $A_{I\to I} = A_{F\to F} = 0$ and the collective is stationary,

$$\begin{aligned}
A_F &= \sum_{j\in\Omega_I} A_{j\to F} \\
&= \sum_{j\in\Omega_I} \frac{|w| F(\theta_I - \theta_F + \pi)}{n_F + |w| n_I} \\
&= \frac{|w| n_I F(\theta_I - \theta_F + \pi)}{n_F + |w| n_I} \\
&\approx \frac{|w|(\pi - \Delta\theta)}{1 + |w|} \quad \text{(D14)}
\end{aligned}$$

and

$$\begin{aligned}
A_I &= \sum_{j\in\Omega_F} A_{j\to I} \\
&= \sum_{j\in\Omega_F} \frac{F(\theta_F - \theta_I)}{n_F + n_I} \\
&= \frac{n_F F(\theta_F - \theta_I)}{n_F + n_I} \\
&= \frac{F(\Delta\theta)}{2} \approx \frac{\Delta\theta}{2}. \quad \text{(D15)}
\end{aligned}$$

At stationary state $\theta_I(t + \Delta t) - \theta_I(t) = \theta_F(t + \Delta t) - \theta_F(t)$ which means

$$\frac{|w|(\pi - \Delta\theta)}{1 + |w|} = \frac{\Delta\theta}{2} \Longrightarrow \Delta\theta = \frac{2|w|}{1 + 3|w|}\pi. \quad \text{(D16)}$$

Then,

$$\begin{aligned}
\langle A_i(t)\rangle_{R_I} &= \sum_{j\in\Omega_I} \frac{|w| F\left(\pi - \Delta\theta + \beta_i + \beta_j\right)}{n_F + |w| n_I} \\
&= \sum_{j\in\Omega_I} \frac{|w|}{n_F + |w| n_I} F\left(\frac{1 + |w|}{1 + 3|w|}\pi + \beta_i + \beta_j\right) \quad \text{(D17)}
\end{aligned}$$

in which $\frac{1+|w|}{1+3|w|}\pi \in \left(\frac{\pi}{3}, \pi\right]$, $\beta_i + \beta_j \in [-\eta, \eta]$, $n_F$ represents the number of neighbor followers, and $n_I$ represents the number of neighbor influencers. In the following calculation, we assume $n_I = n_F$. Denote $W = \frac{1+|w|}{1+3|w|} \in \left(\frac{1}{3}, 1\right]$.

When $-\pi < W\pi + \beta_i + \beta_j < \pi$, one has $-\pi - W\pi < \beta_i + \beta_j < \pi - W\pi$. As $\beta_i$ and $\beta_j$ are symmetric, we have $W\pi - \pi < \beta_i + \beta_j < \pi - W\pi$ which indicates $0 \leq \eta < \pi - W\pi$. Then,

$$\begin{aligned}\langle A_i(t)\rangle_{R_I} &= \sum_{j\in\Omega_I} \frac{|w|}{n_F + |w|n_I} F\big(W\pi + \beta_i + \beta_j\big) \\ &= \sum_{j\in\Omega_I} \frac{|w|}{n_F + |w|n_I} \big(W\pi + \beta_i + \beta_j\big) \\ &\approx \frac{n_I|w|}{n_F + |w|n_I}(W\pi + \beta_i) \\ &\approx \frac{|w|}{1+|w|}(W\pi + \beta_i). \end{aligned} \tag{D18}$$

As $W\pi + \beta_i \in \left[-\frac{\eta}{2} + W\pi, \frac{\eta}{2} + W\pi\right]$ and $\eta < \pi - W\pi$,

$$-\frac{\eta}{2} + W\pi > \frac{3}{2}W\pi - \frac{\pi}{2} > 0. \tag{D19}$$

Thus $W\pi + \beta_i > 0$,

$$\begin{aligned}\left|\langle A_i(t)\rangle_{R_I}\right| &= \left|\frac{|w|}{1+|w|}(W\pi + \beta_i)\right| \\ &= \frac{|w|}{1+|w|}(W\pi + \beta_i) \end{aligned} \tag{D20}$$

and

$$\begin{aligned}\langle|\langle A\rangle_{R_I}|\rangle_{N_F,T} &= \frac{|w|}{1+|w|}W\pi \\ &= \frac{|w|}{1+|w|}\frac{1+|w|}{1+3|w|}\pi \\ &= \frac{|w|}{1+3|w|}\pi. \end{aligned} \tag{D21}$$

When $\eta > \pi - W\pi$, $F\big(W\pi + \beta_i + \beta_j\big)$ becomes complicated and the assumptions we have made above become invalid. However, we can obtain $\langle|\langle A\rangle_{R_I}|\rangle_{N_F,T}$ when $\eta = 2\pi$. Using the central limit theory,

$$\begin{aligned}\langle A_i(t)\rangle_{R_I} &= \sum_{j\in\Omega_I} \frac{|w|}{n_F + |w|n_I} U \\ &= \frac{n_I|w|}{n_F + n_I|w|} \sum_{j\in\Omega_I} \frac{U}{n_I} \\ &\sim \frac{n_I|w|}{n_F + |w|n_I} \mathcal{N}\left(0, \frac{\pi^2}{3n_I}\right). \end{aligned} \tag{D22}$$

$U$ is uniformly distributed among $(-\pi, \pi]$. Thus,

$$\langle|\langle A\rangle_{R_I}|\rangle_{N_F,T} = \frac{|w|}{1+|w|}\sqrt{\frac{2}{3R^2\rho_I}}. \tag{D23}$$

When $w = -1$, $R = 5$, and $\rho_I = 2$, $\langle|\langle A\rangle_{R_I}|\rangle_{N_F,T} = 0.058$ for $\eta = 2\pi$ which is consistent with the simulation results.

**Case 3**: Transition between chiral and aligned phase

We already know that the averaged influence the influencers give to the followers is

$$\langle A_i(t)\rangle_{R_I} = \sum_{j\in\Omega_I} \frac{|w|}{n_F + n_I|w|} F\big(W\pi + \beta_i + \beta_j\big). \tag{D24}$$

For $w < 0$ and $|w| > 1$, $\frac{1}{3}\pi < \frac{1+|w|}{1+3|w|}\pi < \frac{1}{2}\pi$. When $\eta < \frac{1}{2}\pi$ and varying $w$,

$$\left|\frac{1+|w|}{1+3|w|}\pi + \beta_i + \beta_j\right| < \pi. \tag{D25}$$

Therefore,

$$\begin{aligned}\langle A_i(t)\rangle_{R_I} &= \sum_{j\in\Omega_I} \frac{|w|}{n_F + n_I|w|} \big(W\pi + \beta_i + \beta_j\big) \\ &= \sum_{j\in\Omega_I} \frac{|w|}{n_F + n_I|w|}(W\pi + \beta_i) \\ &= \frac{n_I|w|}{n_F + n_I|w|}(W\pi + \beta_i) \\ &= \frac{|w|}{1+|w|}(W\pi + \beta_i). \end{aligned} \tag{D26}$$

As $\eta < \frac{1}{2}\pi$, $|\beta_i| < \frac{1}{4}\pi$ and $\left(\frac{1+|w|}{1+3|w|}\pi + \beta_i\right) > 0$. Thus,

$$\left|\langle A_i(t)\rangle_{R_I}\right| = \frac{|w|}{1+|w|}(W\pi + \beta_i) \tag{D27}$$

and

$$\langle|\langle A\rangle_{R_I}|\rangle_{N_F,T} = \frac{|w|}{1+3|w|}\pi. \tag{D28}$$

When $w > 0$, similar to the analysis of the transition between aligned and disordered phase,

$$\langle A_i(t)\rangle_{R_I} = \sum_{j\in\Omega_I} \frac{|w|}{n_F + n_I|w|} F\big(\beta_i + \beta_j\big). \tag{D29}$$

When $\eta < \pi$ and varying $w$, $\left|\beta_i + \beta_j\right| < \pi$,

$$\begin{aligned}\langle A_i(t)\rangle_{R_I} &= \sum_{j\in\Omega_I} \frac{|w|}{n_F + n_I|w|} \big(\beta_i + \beta_j\big) \\ &= \frac{n_I|w|}{n_F + n_I|w|}\beta_i = \frac{|w|}{1+|w|}\beta_i \end{aligned} \tag{D30}$$

and

$$\left|\langle A_i(t)\rangle_{R_I}\right| = \frac{|w|}{1+|w|}|\beta_i|. \tag{D31}$$

Where $|\beta_i|$ is uniformly distributed in $\left[0, \frac{\eta}{2}\right]$. Therefore,

$$\langle |\langle A \rangle_{R_I}| \rangle_{N_F,T} = \frac{|w|}{1+|w|}\frac{\eta}{4}. \quad \text{(D32)}$$

## APPENDIX E: THREE BINARY CASES FOR DIFFERENT PID METHODS

The XOR case [Table I] is the typical example of purely synergistic information. Consequently, all the PID methods correctly attribute 1 bit to $Syn$ with 0 $Shd$ and $Unq$ components, as shown in Table II.

TABLE I. Operation XOR.

| $S_1$ | $S_2$ | $T$ | $Pr$ |
|---|---|---|---|
| 0 | 0 | 0 | 1/4 |
| 0 | 1 | 1 | 1/4 |
| 1 | 0 | 1 | 1/4 |
| 1 | 1 | 0 | 1/4 |

For the OR cases [Table III], we expect all four PID components—$Unq(S_1)$, $Unq(S_2)$, $Shd$, and $Syn$ —to be non-zero. This expectation follows because either source alone can determine the target: if $S_1$ is 1, then target $T = 1$ regardless of $S_2$. Meanwhile, $S_1$ and $S_2$ can both be one simultaneously, causing the same outcome. Thus, we think there is shared information. However, shared information cannot account for all the TE, as $S_1$ and $S_2$ are independent, meaning they are not commonly driven or synchronized [16]. Therefore, $Unq(S_1)$ and $Unq(S_2)$ are supposed to be non-zero. Moreover, knowing $S_1 = 0$, the other source should be known to determine $T$, indicating the existence of the synergistic information. Among the tested methods, $I_{ccs}$ and $I_{dep}$ yield decompositions that align with these qualitative requirements, while $I_{imi}$ is a close second [Table IV]. We note, however, that $I_{dep}$ is computationally expensive [Table V], which precludes its use in our large–scale numerical studies.

For the ADD case [Table VI], we expect $Unq(S_1)$, $Unq(S_2)$, and $Syn$ to be non-zero. If $S_1 = 0$, the target $T$ is restricted to $\{0, 1\}$; if $S_1 = 1$, the target $T$ is restricted to $\{1, 2\}$. The same holds for $S_2$. Thus, each source provides some unique constraint on $T$. However, knowing only $S_1$ or only $S_2$ is insufficient to determine $T$ exactly; both must be combined to resolve the uncertainty of $T$, which satisfies the definition of synergistic information. Thus, ADD contains unique and synergistic information. Among the tested methods, $I_{ccs}$ and $I_{imi}$ yield decompositions consistent with this expectation [Table VII].

The difference between OR and ADD is that, for OR, knowing $S_1 = 1$ or $S_2 = 1$, is sufficient to determine $T = 1$; Thus, the configuration $(S_1, S_2, T) = (1,1,1)$ contributes to shared information. In contrast, for the ADD case, knowing only that $S_1 = 1$ or $S_2 = 1$ leaves $T$ uncertain; the outcome is fully resolved only when both sources are known. Consequently, the configurations, $(S_1, S_2, T) = (1,1,2)$ and $(0,0,0)$, contribute to the synergistic information.

Based on the above analysis, we decide to choose $I_{css}$ and $I_{imi}$ as representative methods for our detailed collective analysis. To provide comparison, we also examine the historically first PID method, $I_{min}$ [33].

To estimate the information transfer in a collective system, we first compute the information from neighbors to a particle and then average this quantity over a subset of particles; this sample average serves as our estimate of the system-wide information transfer. We investigated how many particles must be included to obtain a reliable average. Using parameters $N = 1000$, $w_{i\to j} = 1$, $R = 1$, and $\rho = 4$,

TABLE II. The results of different PID methods for the XOR operation.

| | $I_{min}$ | $I_{proj}$ | $I_{broja}$ | $I_{ccs}$ | $I_{pm}$ | $I_{dep}$ | $I_{imi}$ |
|---|---|---|---|---|---|---|---|
| $Unq(S_1)$ | 0 | 0 | 0 | 0 | 0 | 0 | 0 |
| $Unq(S_2)$ | 0 | 0 | 0 | 0 | 0 | 0 | 0 |
| $Shd$ | 0 | 0 | 0 | 0 | 0 | 0 | 0 |
| $Syn$ | 1 | 1 | 1 | 1 | 1 | 1 | 1 |

TABLE III. Operation OR.

| $S_1$ | $S_2$ | $T$ | $Pr$ |
|---|---|---|---|
| 0 | 0 | 0 | 1/4 |
| 0 | 1 | 1 | 1/4 |
| 1 | 0 | 1 | 1/4 |
| 1 | 1 | 1 | 1/4 |

TABLE IV. The results of different PID methods for the OR operation.

| | $I_{min}$ | $I_{proj}$ | $I_{broja}$ | $I_{ccs}$ | $I_{pm}$ | $I_{dep}$ | $I_{imi}$ |
|---|---|---|---|---|---|---|---|
| $Unq(S_1)$ | 0 | 0 | 0 | 0.208 | −0.25 | 0.230 | 0.311 |
| $Unq(S_2)$ | 0 | 0 | 0 | 0.208 | −0.25 | 0.230 | 0.311 |
| $Shd$ | 0.311 | 0.311 | 0.311 | 0.104 | 0.561 | 0.082 | 0 |
| $Syn$ | 0.5 | 0.5 | 0.5 | 0.292 | 0.75 | 0.270 | 0.189 |

TABLE V. Time consumption for different PID methods. The time is averaged using 5 computing with $\eta = \pi$, $N = 400$, $\rho = 4$, $w_{i\to j} = 1$, and 8 bins.

| | $I_{min}$ | $I_{proj}$ | $I_{broja}$ | $I_{ccs}$ | $I_{pm}$ | $I_{dep}$ | $I_{imi}$ |
|---|---|---|---|---|---|---|---|
| **Time (s)** | 0.69 $\pm$ 0.02 | 0.25 $\pm$ 0.02 | 905.74 $\pm$ 22.68 | 539.43 $\pm$ 25.92 | 0.19 $\pm$ 0.01 | $> 2000$ | 4.43 $\pm$ 0.36 |

TABLE VI. Operation ADD.

| $S_1$ | $S_2$ | $T$ | $Pr$ |
|---|---|---|---|
| 0 | 0 | 0 | 1/4 |
| 0 | 1 | 1 | 1/4 |
| 1 | 0 | 1 | 1/4 |
| 1 | 1 | 2 | 1/4 |

TABLE VII. The results of different PID methods for the ADD operation.

| | $I_{min}$ | $I_{proj}$ | $I_{broja}$ | $I_{ccs}$ | $I_{pm}$ | $I_{dep}$ | $I_{imi}$ |
|---|---|---|---|---|---|---|---|
| $Unq(S_1)$ | 0 | 0 | 0 | 0.5 | 0 | 0.311 | 0.5 |
| $Unq(S_2)$ | 0 | 0 | 0 | 0.5 | 0 | 0.311 | 0.5 |
| $Shd$ | 0.5 | 0.5 | 0.5 | 0 | 0.5 | 0.189 | 0 |
| $Syn$ | 1 | 1 | 1 | 0.5 | 1 | 0.689 | 0.5 |

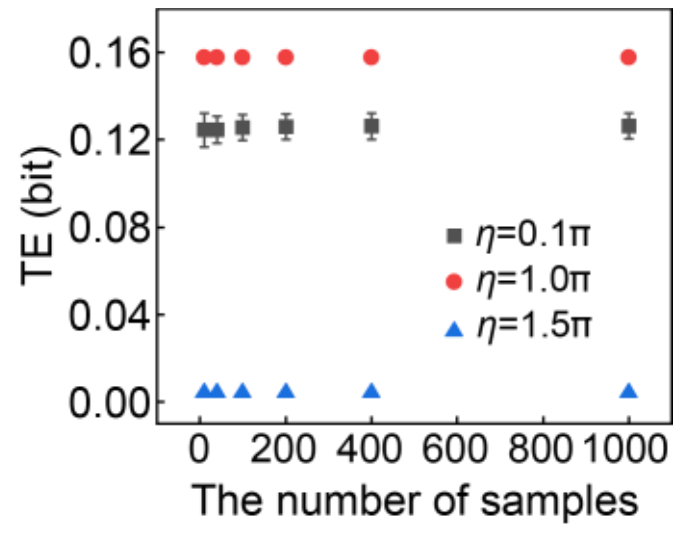

FIG. 20. TE versus the number of samples ($\{10, 40, 100, 200, 400, 1000\}$) for different $\eta$ with $R = 1, N = 1000$ and $\rho = 4$. It shows that the value of TE is insensitive to the number of samples in this range of 10 – 1000.

we computed TE as a reference for noise strengths $\eta = 0.1\pi, 1.0\pi, 1.5\pi$ while varying the sample size. Fig. 20 shows that the average value is not sensitive to the number of particles selected from 10 to 1000. Therefore, to balance computational cost with statistical reliability, we routinely sample 100 particles for averaging. An exception is made for the computationally demanding $I_{ccs}$ method, for which we use only 10 particles [Table V].

## APPENDIX F: LIST OF SYMBOLS

The following list summarizes the symbols used in this paper in the order they appear. All mathematical variables are italicized; vectors are set in bold.

|  |  |
|---|---|
| $X, Y$ | Capitalized letters represent random variable |
| $x, y$ | Small letters represent the specific value of a random variable |
| $\mathcal{X}, \mathcal{Y}$ | Handwritten capital letters represent alphabets |
| $H(\,)$ | Shannon entropy |
| $p(\,)$ | Probability mass function |
| $f(\cdot)$ | Probability density function |
| $D(\,)$ | Kullback-Leibler distance |
| $I(\,)$ | Mutual information |
| $t$ | time |
| $\Delta t$ | Discrete time step |

$TDMI_{X\to Y}$ Time-delayed mutual information from $X$ to $Y$

$TE_{X\to Y}$ Transfer entropy from $X$ to $Y$

$Unq(\cdot)$ Unique information or intrinsic information

$Shd$ Shared information or redundant information

$Syn$ Synergistic information

$\theta_i$ Orientation of particle $i$

$\beta_i$ Noise of particle $i$

$\eta_i$ Noise strength of particle $i$

$F(\cdot)$ Wrapping function that maps angular differences to $(-\pi, \pi]$

$w_{j\to i}$ Interaction strength particle $j$ exerts on particle $i$

$\Theta(\cdot)$ Heaviside step function

$R$ Neighborhood cutoff radius

$A_{j\to i}$ Pairwise instantaneous influence of particle $j$ on particle $i$

$A_i$ Total influence on particle $i$ from all neighbors

$v$ Speed of a particle

$\boldsymbol{r}_i$ Position vector

$\boldsymbol{v}_i$ Velocity vector

$L$ Size of the arena

$\rho$ Particle density

$\Omega_I$ , $\Omega_F$ Index sets for the influencers and the followers

$v_a$ Mean speed

$\chi$ Susceptibility

$G$ Binder cumulant

$T$ Number of times steps

$N_F$ Number of followers

$\langle\cdot\rangle_{R_I}$ Neighbor-average operation over influencer-neighbor within the cutoff radius

$\langle\cdot\rangle_{N_F}$ Number-average operation over followers

$\langle\cdot\rangle_T$ Time-average operation

$|\cdot|$ Norm operation

$n_I, n_F$ Number of influencer-neighbor and follower-neighbor

$\theta_{nbs}$ Aggregated set of influencer-neighbors

$I_{min}$ PID method using minimum specific information

$I_{ccs}$ PID method based on pointwise changes in surprisal

$I_{imi}$ PID method using intrinsic mutual information